\documentclass[10pt, twosided, twocolumn, conference]{IEEEtran}
\usepackage{epsfig}
\usepackage{latexsym}
\usepackage{bbold,bbm,dsfont}
\usepackage{amsmath}
\newtheorem{theorem}{Theorem}[section]
\newtheorem{lemma}{Lemma}[section]

\newtheorem{remarks}{Remarks}[section]
\newtheorem{corollary}{Corollary}[section]

\title{Optimal Routing and Power Control
    for a \mbox{Single Cell}, Dense, \mbox{Ad Hoc} Wireless Network}

\author{Venkatesh~Ramaiyan, Anurag~Kumar,~\IEEEmembership{Fellow,~IEEE,} and Eitan
    Altman,~\IEEEmembership{Member,~IEEE}%
}

\date{}
\begin{document}
\maketitle
\thispagestyle{empty}
\pagestyle{empty}
\begin{abstract}
  We consider a dense, ad hoc wireless network, confined to a small
  region.
  The wireless network is operated as a single cell, i.e., only one successful
transmission is supported at a time.
  Data packets are sent between source-destination pairs by
  multihop relaying. We assume that nodes
  self-organise into a multihop network such that all hops are of
  length $d$ meters, where $d$ is a design parameter.  There is a
  contention based multiaccess scheme, and it is assumed that every
  node always has data to send, either originated from it or a transit
  packet (saturation assumption).  In this scenario, we seek to
  maximize a measure of the transport capacity of the network
  (measured in bit-meters per second) over power controls (in a fading
  environment) and over the hop distance $d$, subject to an average
  power constraint.

  We first argue that for a dense collection of nodes confined to a
  small region, single cell operation is efficient for single user
  decoding transceivers.  Then, operating the dense ad hoc wireless network
  (described above) as a single cell, we study the hop length
  and power control that maximizes the transport capacity for a given
  network power constraint.  More specifically, for a fading channel
  and for a fixed transmission time strategy (akin to the IEEE 802.11
  TXOP), we find that there exists an intrinsic aggregate bit rate
  ($\Theta_{opt}$ bits per second, depending on the contention
  mechanism and the channel fading characteristics) carried by the
  network, when operating at the optimal hop length and power control.
  The optimal transport capacity is of the form $d_{opt}(\bar{P_t})
  \times \Theta_{opt}$ with $d_{opt}$ scaling as
  $\bar{P_t}^{\frac{1}{\eta}}$, where $\bar{P_t}$ is the available
  time average transmit power and $\eta$ is the path loss exponent.
  Under certain conditions on the fading distribution, we then provide
  a simple characterisation of the optimal operating point.
\end{abstract}


\section{Introduction}

We consider a scenario in which there is a large number of stationary nodes
(e.g., hundreds of nodes) confined to a small area
(e.g., spatial diameter 30m), and organised into a multihop
ad hoc wireless network.
We assume that, traffic in the network is homogeneous
and  data packets are sent between source-destination
pairs by multihop relaying with single user decoding and forwarding of packets,
i.e., signals received from nodes other than the intended
transmitter are treated as interference.
A distributed multiaccess contention scheme is used in
order to schedule transmissions; for
example, the CSMA/CA based distributed coordination function (DCF) of
the IEEE~802.11 standard for wireless local area networks (WLANs).  We
assume that all nodes can decode all the contention control
transmissions (i.e., there are no hidden nodes), and only one
successful transmission takes place at any time in the network.  In
this sense we say that we are dealing with a \emph{single cell}
scenario.
We further assume that, during
the exchange of contention control packets, pairs of communicating
nodes are able to estimate the channel fade between them and
are thus able to perform power control per transmission.

There is a natural tradeoff between using high power and long hop
lengths (single hop direct transmission between the source-destination
pair), versus using low power and shorter hop lengths
(multihop communication using intermediate nodes), with the latter
necessitating more packets to be transported in the network.  The
objective of the present paper is to study optimal routing, in terms
of the hop length, and optimal power
control for a fading channel, when the network (described above)
is used in a multihop mode.  Our
objective is to maximise a certain measure of network transport
capacity (measured in bit-meters per
second; see Section~\ref{sec:multihop_obj}), subject to a network power constraint.
A network power constraint determines, to a
first order, the lifetime of the network.

Situations and considerations such as those that we study could arise
in a dense ad hoc wireless sensor network.  Ad hoc wireless sensor networks are now
being studied as possible replacements for wired measurement networks
in large factories. For example, a distillation column in a chemical
plant could be equipped with pressure and temperature sensors and
valve actuators.  The sensors monitor the system and communicate the
pressure and temperature values to a central controller which in turn
actuates the valves to operate the column at the desired operating
point. Direct communication between the sensors and actuators is also
a possibility. Such installations could involve hundreds of devices,
organised into a single cell ad hoc wireless network because
of the physical proximity of the nodes.  There would be
many flows within the network and there would be multihopping. We wish
to address the question of optimal organisation of such an ad hoc
network so as to maximise its transport capacity subject to a power
constraint. The power constraint relates to the network life-time
and would depend on the application.  In
a factory situation, it is possible that power could be supplied to
the devices, hence large power would be available. In certain
emergencies, ``transient'' sensor networks could be deployed for
situation management; we use the term ``transient'' as these
networks are supposed to exist for only several
minutes or hours, and the devices could be disposable.
Such networks need to have large throughputs,
but, being transient networks, the power constraint could again be loose. On the
other hand sensor networks deployed for monitoring some phenomenon in
a remote area would have to work with very small amounts of power, while sacrificing
transport capacity. Our formulation aims at providing insights into
optimal network operation in a range of such scenarios.

\subsection{Preview of Contributions}
We motivate the definition of the
transport capacity of the network as the product of the aggregate
throughput (in bits per second) and the hop distance (in meters). For random
spatio-temporal fading, we seek the power control and the hop distance
that jointly maximizes the transport capacity, subject to a network
average power constraint.  For a fixed data transmission time strategy
(discussed in Section~\ref{sec:fixed_trans_time}),
we show that the optimal power allocation function has a water pouring form
(Section~\ref{sec:optimize_ph}).
At
the optimal operating point (hop distance and power control) the
network throughput ($\Theta_{opt}$, in bits per second) is
shown to be a fixed quantity, depending only on
the contention mechanism and fading model, but independent of the
network power constraint (Section~\ref{sec:optimize_d}).
Further, we show that the optimal transport capacity is of
the form $d_{opt}(\bar{P_t}) \times \Theta_{opt}$,
with $d_{opt}$ scaling as $\bar{P_t}^{\frac{1}{\eta}}$,
where $\bar{P}_t$ is the
available time average transmit power, and $\eta$ is the power law path loss
exponent (Theorem~\ref{thm:fading_time}). Finally, we provide a
condition on the fading density that leads to a
simple characterisation of the optimal hop distance (Section~\ref{sec:characterise_d}).

\section{Motivation for Single Cell Operation}
\label{sec:optimal_single_cell}
In this context (a dense, ad hoc wireless network),
the seminal paper by Gupta and Kumar \cite{wireless.gupta-kumar00capacity-wireless}
would suggest that each node should communicate with neighbours as close as possible
while maintaining network connectivity. This maximises network transport capacity
(in bit-meters per second), while minimising network average power. It has been
observed by Dousse and Thiran in \cite{wireless.dousse-thiran04connectivity-capacity},
that if, unlike \cite{wireless.gupta-kumar00capacity-wireless},
a practical model of bounded received power for finite transmitter power is used,
then the increasing interference with an increasing density of simultaneous transmitters
is not consistent with a minimum SINR requirement at each receiver.
The following discussion motivates
single cell operation for our framework.

Consider a planar wireless network with $n$ nodes in a square of fixed area $A$.
Let $K(n)$ be the spatial reuse in the network (the number of simultaneous
transmissions) and $r(K(n))$, the maximum transmitter-receiver
separation.
Denote by $P(K(n))$ the common power per transmitter
(assumed to be the same for all nodes)
satisfying a network average power constraint $\bar{P}$
(as in Section~\ref{sec:fading_time}).
The maximum SINR achievable per link in such a network (with single
user decoding receivers) is bounded above by
$\frac{P(K(n))}{(N + I_{K(n)})}$,
where $N$ is the receiver noise power and
$I_{K(n)}$, the interference at a node due to spatial reuse.
Using the finite (and fixed) area assumption, the minimum
interference power from any simultaneous transmission
is given by $\frac{P(K(n))}{(\sqrt{2 A})^{\eta}}$.
Hence, the SINR achievable over any link is bounded above by
$\frac{P(K(n))}{N + (K(n) - 1) \frac{P(K(n))}{(2 A)^{\frac{\eta}{2}}}}$
and the maximum bit rate achievable over a link is,
$\log\left(1 + \frac{P(K(n))}{N + (K(n) - 1) \frac{P(K(n))}{(2 A)^{\frac{\eta}{2}}}}\right)$.
The aggregate throughput in the network is now bounded above by
$C(K(n))$,
\[ C(K(n)) := K(n) \log\left(1 + \frac{P(K(n))}{N + (K(n) - 1) \frac{P(K(n))}{(2 A)^{\frac{\eta}{2}}}}\right) \]
Clearly, $C(K(n))$ is uniformly bounded above\footnote{Note that $C(K(n)) \leq K(n) \log\left( 1 + \frac{(2 A)^{\frac{\eta}{2}}}{(K(n) - 1)} \right)$, independent of the transmit power $P(K(n))$.}.
Also, $r(K(n)) \leq \sqrt{2 A}$. Hence, we see that the transport
capacity
achievable in the network (bounded above by $\sup_{\{K(n) : K(n) \geq 2\}} C(K(n)) r(K(n))$) is finite,
 independent of the number of nodes and the network power $\bar{P}$.
Further, we would expect the
transmitter-receiver separation (bounded above by $r(K(n))$)
to decrease to $0$ as $K(n)$ increases to $\infty$ (finite area assumption).
Hence, $ \lim_{K(n) \rightarrow \infty} C(K(n)) r(K(n)) = 0 $.
This implies that there
exists an optimal $K(n), 1 < K(n) < \infty$,
which maximises the transport
capacity in the network,
i.e., the optimum spatial reuse is finite.
Now, consider a simple TDMA scheme, without spatial reuse and with direct transmission
between the source
and the destination. For a transmit power $\bar{P}$,
the TDMA schedule achieves $\Theta(\log(\bar{P}))$ transport capacity,
i.e., an unbounded transport capacity as a function of the network power $\bar{P}$.
As discussed above,
with spatial reuse, the system becomes interference limited, and hence, becomes
inefficient for large $\bar{P}$.
More recently, El Gamal and Mammen \cite{winet.elgamal-mammen06optimal-hopping}
have shown that, if the transceiver energy and communication overheads at each hop is factored
in, then the operating regime studied in \cite{wireless.gupta-kumar00capacity-wireless}
is neither energy efficient nor delay optimal.
Fewer hops between the transmitter and receiver
(and hence, less spatial reuse) reduce the overhead energy
consumption and lead to a better throughput-delay tradeoff.

While optimal operation of the network might suggest using some
spatial reuse (finite, as discussed above),
coordinating simultaneous transmissions
(in a distributed fashion), in a constrained area, is extremely difficult
and the associated time, energy and synchronisation overheads have to
be accounted for.
In view of the above discussion, in this paper, we assume that the medium
access control (MAC) is such that only one transmitter-receiver pair
communicate at any time in the network.

\subsection{Outline of the Paper}
In Section~\ref{sec:model_def} we describe the system model and in
Section~\ref{sec:multihop_obj} we motivate the objective.
We study the transport capacity of a single cell multihop wireless network, operating in the
fixed transmission time mode, in Section~\ref{sec:fading_time}.
Section~\ref{sec:conclusion} concludes the paper
and discusses future work.

\section{The Network  Model}
\label{sec:model_def}
There is a dense collection of immobile nodes that use multiaccess
multihop radio communication with single user decoding and packet
forwarding to transport packets between various
source-destination pairs.
\begin{itemize}
\item All nodes use
the same contention mechanism with the same parameters (e.g., all
nodes use IEEE~802.11 DCF with the same back-off parameters).
\item  We assume that nodes send control packets
(such as RTS/CTS in IEEE 802.11) with a constant power (i.e., power
control is not used for the control packets) during contention, and
these control packets are decodable by \emph{every} node in the
network.  As in IEEE~802.11, this can be done by using a low rate,
robust modulation scheme and by restricting the diameter of the
network.  This is the ``single cell'' assumption, also used in
\cite{wanet.kumar-etal04new-insights}, and implies that there can be
only one successful ongoing transmission at any time.
\item During the control packet exchange, each transmitter learns about the
channel ``gain'' to its intended receiver, and decides upon the power
level that is used to transmit its data packet.  For example, in
IEEE 802.11, the channel gain to the
intended receiver could be estimated during the RTS/CTS control packet
exchange.  Such channel information can then be used by the
transmitter to do power control.
In our paper, we assume that such channel estimation and power control
is possible on a transmission-by-transmission basis.
\item In this work, we
model only an average power constraint and not a peak power constraint.
\item Saturation assumption : We assume that the traffic is homogeneous in the network
and all the nodes have data to send at all times; these
could be locally generated packets or transit packets.
In \cite{wireless.hyytia-virtamo06load-balancing},
the authors study the problem of load balancing in dense multihop wireless
networks with arbitrary traffic requirements.
In our work, we do not restrict to straight line paths,
and permit such a load balancing routing strategy as in \cite{wireless.hyytia-virtamo06load-balancing},
which ensures that the load and the
channel access pattern are identical for all the nodes.
\end{itemize}
Data packets are sent between source-destination pairs by multihop relaying.
Based on the dense network and traffic homogeneity assumption, we further make the following
assumption.
\begin{itemize}
\item The nodes self-organise so that all hops are of length $d$,
  i.e., a one hop transmission always traverses a distance of
  $d$~meters. This hop distance, $d$, will be one of our optimisation
  variables.
\end{itemize}
For a random node deployment, the hop distance
  that maximizes the system throughput need not be the same for every node
  and every flow. However, the approximation holds good for a
  homogeneous network with large number of nodes. Further, it will be practically infeasible
  to optimize every hop in a dense setup with hundreds of nodes.

\subsection{Channel Model: Path Loss, Fading and Transmission Rate}
\label{sec:channel_model}
The channel gain between a transmitter-receiver pair for a hop is
assumed to be a
function of the hop length ($d$) and the multipath fading ``gain''
($h$).
The path loss for a hop distance $d$ is given by $\frac{1}{d^{\eta}}$,
where $\eta$ is the path loss exponent, chosen depending on the
propagation characteristics of the environment (see, for e.g.,
\cite{book.rappaport.wireless-communications}). This variation of path
loss with $d$ holds for $d > d_0$, the far field reference distance;
we will assume that this inequality holds (i.e., $d > d_0$), and will justify this
assumption in the course of the analysis (see Theorem~\ref{thm:fading_time}).

We assume a flat and slow fading channel with additive
white Gaussian noise of power $\sigma^2$.
We assume that for each transmitter-receiver pair, the channel gain
due to multipath fading may change from transmission to transmission, but
remains constant over any packet transmission duration. Since
successive transmissions can take place between randomly selected
pairs of nodes (as per the outcome of the distributed contention
mechanism) we are actually modeling a spatio-temporal fading process.
We assume that this fading process is stationary in space and time
with some given marginal distribution $H$.
Let the cumulative distribution of $H$ be $A(h)$ (with a p.d.f. $a(h)$), which by our
assumption of spatio-temporal stationarity of fading is the same for
all transmitter-receiver pairs and for all transmissions.
We assume that the channel coherence time, $\tau_c$, applicable to all the links in the network,
upper bounds every data transmission duration in the network.
Further, we assume that $H$ and $\tau_c$ are independent of the hop distance $d$.

When a
node transmits to another node at a distance $d$ (in the transmitting
antenna's far field), using transmitter power $P$, with channel power
gain due to fading, $h$, then we assume that the transmission rate
given by Shannon's formula is achieved over the transmission burst;
i.e., the transmission rate is given by
\begin{eqnarray*}
  C = W \log \left(1 + \frac{hP \alpha}{\sigma^2 d^\eta} \right)
\end{eqnarray*}
where $W$ is the signal bandwidth and $\alpha$
is a constant accounting for any fixed power gains between
the transmitter and the receiver.
Note that this requires that the transmitter has available several
coding schemes of different rates, one of which is chosen for each
channel state and power level.

\subsection{Fixed Transmission Time Strategy}
\label{sec:fixed_trans_time}
We consider a fixed transmission time scheme,
where all data transmissions are of equal duration, $T$ ($< \tau_c$) secs,
independent of the bit rate achieved over the wireless link.
This implies that the amount of data that a transmitter
sends during a transmission opportunity is proportional to the
achieved physical link rate.
Upon a successful control
packet exchange, the channel (between the transmitter, that ``won''
the contention, and its intended receiver) is reserved for a duration
of $T$ seconds independent of the channel state $h$. This is akin to the
``TxOP'' (transmission opportunity) mechanism in the IEEE~802.11 standard.
Thus, when the power allocated during the channel state $h$ is  $P(h)$,
$C(h) T$ bits are sent across the channel, where
$ C(h) = W \log\left( 1 + \frac{P(h) h \alpha}{\sigma^2 d^{\eta}} \right) $.
When $P(h) = 0$, we assume that  the channel is left
idle for the next $T$ seconds.
The transmitter does not relinquish the channel immediately, and the channel
reserved for the transmitter-receiver pair (for example, by the RTS/CTS signalling)
is left empty for the duration of $T$ seconds.

The optimality of a fixed transmission time scheme, for throughput, as compared
to a fixed packet length scheme, can be formally established
(see Appendix~\ref{sec:mh_ftt_vs_fp}), we only
provide an intuition here. When using fixed packet lengths,
a transmitter may be forced to send the entire packet even if the
channel is poor, thus taking longer time and more power.
On the other hand, in a fixed transmission time scheme, we send more data
when the channel is good and limit our inefficiency when the channel
is poor.

\section{Multihop Transport Capacity}
\label{sec:multihop_obj}

Let $d$ denote the common hop length and $\{P(h)\}$ a power allocation policy, with $P(h)$
denoting the transmit power used when the channel state is $h$.
We take a simple model for the random access channel contention
process. The channel goes through successive contention periods.
Each period can be either an idle slot, or a collision period, or a
successful transmission with probabilities $p_i, p_c$ and $p_s$
respectively.
Under the node saturation assumption, the aggregate bit rate carried by
the system, $\Theta_T(\{P(h)\},d)$,
for the hop distance $d$ and power allocation $\{P(h)\}$, is given by
(see
\cite{wanet.kumar-etal04new-insights})
\begin{eqnarray}
\Theta_T(\{P(h)\},d):= \frac{p_s (\int_0^\infty L(h)~ {\rm d}A(h)~)}{p_i T_i + p_c T_c + p_s (T_o + T)}
\label{eqn:phi_case2}
\end{eqnarray}
where $L(h) := C(h) T$ (and $C(h)$ is a function of $\{P(h)\}$ and $d$).
 $T_i, T_c$ and $T_o$ are the average
time overheads associated with an idle slot, collision and data transmission.
For example, in IEEE~802.11 with the RTS/CTS mechanism being used,
a collision takes a fixed time independent of the data transmission rate.
We note that $p_i, p_s, p_c, T_i, T_o, \mbox{and} \ T_c$ depend only
on the parameters of the distributed contention mechanism (MAC
protocol) and the channel, and not on any of the decision variables that we
consider.

With $\Theta_T(\{P(h)\},d)$ defined as in (\ref{eqn:phi_case2}),
we consider $\Theta_T(\{P(h)\},d) \times d$ as our measure of transport capacity
of the network. This measure can be motivated in several ways.
$\Theta_T(\{P(h)\},d)$ is the rate at which bits are transmitted by the network
nodes. When transmitted successfully, each bit traverses a distance
$d$. Hence, $\Theta_T(\{P(h)\},d) \times d$ is the rate of spatial progress of the flow
of bits in the network (in bit-meters per second).
Viewed alternatively, it is the weighted average of the end-to-end flow throughput
with respect to the distance traversed.
Suppose that a flow $i$ covers a distance $D_i$ with $\frac{D_i}{d}$ hops
(assumed to be an integer for this argument).
Let $\beta_i \Theta_T(\{P(h)\},d)$ be the fraction of
throughput of the network that belongs to flow $i$.
Then, $\frac{\beta_i \Theta_T(\{P(h)\},d)}{\frac{D_i}{d}}$
is the end-to-end throughput for flow $i$ and $\frac{\beta_i \Theta_T(\{P(h)\},d)}{\frac{D_i}{d}} \times D_i
= \beta_i \Theta_T(\{P(h)\},d) \times d$ is the end-to-end throughput
for flow $i$ in bit-meters per second.
Summing over all the flows, we have $\Theta_T(\{P(h)\},d) \times d$, the
aggregate end-to-end flow throughput in bit-meters per second.

With the above motivation, our aim in this paper is to maximise the
quantity $\Theta_T(\{P(h)\},d) \times d$ over the hop distance $d$ and over the power
control $\{P(h)\}$, subject to a \emph{network average power constraint},
$\bar{P}$.  We use a network power constraint that
accounts for the energy used in data transmission as well as the
energy overheads associated with communication.
The network average power, ${\cal P}(\{P(h)\})$, is given by,
\begin{eqnarray}
\lefteqn{{\cal P}(\{P(h)\}) :=} \nonumber \\
& & \frac{p_i E_i + p_c E_c + p_s (E_o + T \int_0^\infty P(h)~ {\rm d}A(h)~)}{p_i T_i + p_c T_c + p_s (T_o + T)}
\label{eqn:pow_case2}
\end{eqnarray}
$E_i, E_c$ and $E_o$ correspond to the energy overheads associated
with an idle period, collision and successful transmission. Thus, $E_i$
denotes the total energy expended in the network over an idle slot,
$E_c$ denotes the total average energy expended by the colliding nodes, as
well as the idle energy of the idle nodes, and $E_o$ denotes the
average energy expended in the successful contention negotiation between
the successful transmitter-receiver pair, the receive energy at the
receiver (in the radio and in the packet processor), and the idle
energy expended by all the other nodes over the time $T_o + T$.
We assume that $E_i, E_c$ and $E_o$ depend only on  the contention
mechanism and not on the decision variables $d$ and $\{P(h)\}$.

\section{Optimising the Transport Capacity}
\label{sec:fading_time}

For a given $\{P(h)\}$ and $d$, and the corresponding throughput $\Theta_T(\{P(h)\},d)$,
the transport capacity in bit-meters per
second, which we will denote by $\psi(\{P(h)\},d)$, is given by
\[ \psi(\{P(h)\},d) := \Theta_T(\{P(h)\},d) \times d \]

Maximizing $\psi(\cdot,\cdot)$ involves optimizing over
$d$, as well as $\{P(h)\}$. However, we observe that, it would not be
possible to vary $d$ with fading, as routes cannot vary at the fading
time scale. Hence, we propose to optimize first over $\{P(h)\}$ for a
given $d$, and then optimize over $d$, i.e., we seek to solve the
following problem,
\begin{eqnarray}
\label{eqn:optimization_main}
\max_{d} \max_{\{\{P(h)\}: {\cal P}(\{P(h)\}) \leq \bar{P}\}} \psi(\{P(h)\},d)
\end{eqnarray}

For a given $d$ and power allocation $\{P(h)\}$,
define the time average transmission power, $\bar{P}_t(\{P(h)\},d)$, and the
time average overhead power, $\bar{P}_o$, as
\begin{eqnarray*}
\bar{P}_t(\{P(h)\},d) &:=& \frac{p_s (\int_0^\infty P(h)~ {\rm d}A(h)~) T}{p_i T_i + p_c T_c + p_s (T_o + T)} \\
\bar{P}_o &:=& \frac{p_i E_i + p_c E_c + p_s E_o}{p_i T_i + p_c T_c + p_s (T_o + T)} \\
\end{eqnarray*}
Observe that $\bar{P}_o$ does not depend on $\{P(h)\}$ and $d$.
Now, the network power constraint can be viewed as
\begin{eqnarray*}
  \bar{P}_t(\{P(h)\},d) \leq \bar{P} - \bar{P}_o
\end{eqnarray*}
where the right hand side is independent of $\{P(h)\}$ or $d$.
$\bar{P_t} := \bar{P} - \bar{P_o}$, is the \emph{time average}
transmitter power constraint for the network.

\subsection{Optimization over $\{P(h)\}$ for a fixed $d$}
\label{sec:optimize_ph}
Consider the optimization problem (from (\ref{eqn:optimization_main}))
\begin{equation}
\label{eqn:optimization_sub}
\max_{\{\{P(h)\}: {\cal P}(\{P(h)\}) \leq \bar{P}\}} \psi(\{P(h)\},d)
\end{equation}
The denominators of $\Theta_T(\cdot,\cdot)$
in (\ref{eqn:phi_case2}) and of ${\cal P}$
in (\ref{eqn:pow_case2}) are independent of $d$ and the power
control $\{P(h)\}$.  Thus, with $d$ fixed, the optimization problem
simplifies to maximizing $\int_0^{\infty} L(h)~ {\rm d}A(h)$ or,
\begin{eqnarray*}
\int_0^\infty \log\left(1 + \frac{P(h) h \alpha}{\sigma^2 d^{\eta}}\right)~ {\rm d}A(h)
\end{eqnarray*}
subject to the power constraint,
\begin{eqnarray*}
\int_0^\infty P(h)~ {\rm d}A(h) \leq \bar{P_t}^{\prime}
\end{eqnarray*}
where $\bar{P_t}^{\prime}$ is given by,
\[ \bar{P_t}^{\prime} := \frac{(p_i T_i + p_c T_c + p_s (T_o + T))}{p_s T} \bar{P_t} \]
$\bar{P_t}^{\prime}$ is the average transmit power constraint averaged only over the
transmission periods (successful contention slots).

Without a peak power constraint, this is a well-known problem whose
optimal solution has the water-pouring form (see \cite{wireless.goldsmith-varaiya97fading-capacity}).
The optimal power allocation function $\{P(h)\}$ is given by
\begin{eqnarray*}
P(h) = \left(\frac{1}{\lambda} - \frac{d^{\eta} \sigma^2}{h \alpha}\right)^{+}
\end{eqnarray*}
where $\lambda$ is obtained from the power constraint equation
\[ \int_{\frac{\lambda \sigma^2 d^{\eta}}{\alpha}}^{\infty} a(h) P(h) dh = \bar{P_t}^{\prime} \]
The optimal power allocation is a nonrandomized policy, where a node
transmits with power $P(h)$ every time the channel is in state $h$
(whenever $P(h) > 0$), or leaves the channel idle for $h$ such
that $P(h) = 0$.

\subsection{Optimization over $d$}
\label{sec:optimize_d}
By defining $\xi(h) := \frac{P(h)}{d^\eta}$, the problem
of maximising the throughput over power controls, for a fixed $d$, becomes
\begin{eqnarray*}
  \max \int_{0}^{\infty} \log\left(1 + \frac{\alpha h}{\sigma^2} \xi(h)\right) a(h) dh
\end{eqnarray*}
subject to
\begin{eqnarray*}
\int_{0}^{\infty}  \xi(h) a(h) dh \leq \frac{\bar{P_t}^{\prime}}{d^{\eta}}
\end{eqnarray*}
Observe that $\bar{P}_t^{\prime}$ and $d$ influence the optimization
problem only as $\frac{\bar{P}_t^{\prime}}{d^{\eta}}$.
Denoting by $\Gamma\left(\frac{\bar{P_t}^{\prime}}{d^{\eta}}\right)$ the
optimal value of this problem,  the problem of optimisation
over the hop-length, $d$, now becomes
\begin{eqnarray}
\label{eqn:opt_over_d}
 \max_{d} d \times \Gamma\left(\frac{\bar{P_t}^{\prime}}{d^{\eta}}\right)
\end{eqnarray}

\begin{theorem}
  \label{thm:dgamma_properties}
  In the problem defined by (\ref{eqn:opt_over_d}),
  the objective $d \times
  \Gamma\left(\frac{\bar{P_t}^{\prime}}{d^{\eta}}\right) $, when viewed
  as a function of $d$, is continuously differentiable. Further,
  when the channel fading random variable, $H$, has a finite mean ($\mathbf{E}(H) < \infty$), then
\begin{enumerate}
\item  $\lim_{d \to 0} d \times \Gamma\left(\frac{\bar{P_t}^{\prime}}{d^{\eta}}\right) =0$
and,
\item  if in addition, $\eta \geq 2$, $\frac{1}{h^2} a\left(\frac{1}{h}\right)$
  is continuously differentiable and $\mathbf{P}(H > h) = O\left(\frac{1}{h^2}\right)$ for large $h$, then,
  $\lim_{d \to \infty} d \times
  \Gamma\left(\frac{\bar{P_t}^{\prime}}{d^{\eta}}\right) =0$,
\end{enumerate}
\end{theorem}
\begin{proof}
The proofs of continuous differentiability of $d \times \Gamma\left(\frac{\bar{P_t}^{\prime}}{d^{\eta}}\right)$,  1) and 2)
are provided
in Appendix~\ref{sec:dgamma_limits}.
\end{proof}

\begin{remarks}
\end{remarks}
\begin{enumerate}
\item Under the conditions proposed in Theorem~\ref{thm:dgamma_properties},
it follows that $d \times \Gamma\left(\frac{\bar{P_t}^{\prime}}{d^{\eta}}\right)$
is bounded over $d \in [0, \infty)$ and achieves its maximum
in $d \in (0, \infty)$.
\item When the objective function (\ref{eqn:opt_over_d}) is
  unbounded, the optimal solution occurs at $d = \infty$ (follows from the continuity results).
\item We note that, in practice, $\eta \geq 2$.
\end{enumerate}

Let $d_0$ be the far field reference distance (discussed in Section~\ref{sec:channel_model}).
\begin{theorem}
\label{thm:fading_time}
  The following hold for the problem in (\ref{eqn:opt_over_d}),
\begin{enumerate}
\item Without the constraint $d > d_0$, the optimum hop distance
  $d_{opt}$ scales as $(\bar{P_t}^\prime)^{\frac{1}{\eta}}$.
\item There is a value $\bar{P_t}^\prime_{\min}$ such that, for $\bar{P_t}^{\prime}
  > \bar{P_t}^\prime_{\min}$, $d_{opt} > d_0$, and the optimal
  solution obeys the scaling shown in 1).
\item For $\bar{P_t}^{\prime} > \bar{P_t}^{\prime}_{\min}$, the optimum power
	control $\{P(h)\}$ is of the water pouring
  form and scales as $\bar{P_t}^\prime$.
\item For $\bar{P_t}^{\prime} > \bar{P_t}^{\prime}_{\min}$, the optimal transport capacity scales as
  $(\bar{P_t}^\prime)^{\frac{1}{\eta}}$.
\end{enumerate}
\end{theorem}
\begin{proof}
  \begin{enumerate}
  \item Let $d_{opt}$ be optimal for $\bar{P_t}^\prime > 0$.  We claim
    that, for $x>0$, $x^{\frac{1}{\eta}} d_{opt}$ is optimal for the
    power constraint $x \bar{P_t}^\prime$. For suppose this was not so,
    it would mean that there exists $d>0$ such that
    \begin{eqnarray*}
      \left( x^{\frac{1}{\eta}} d_{opt} \
        \Gamma\left(\frac{x \bar{P_t}^\prime}{(x^{\frac{1}{\eta}} d_{opt})^{\eta}} \right) \right)<  d \ \Gamma\left(\frac{x \bar{P_t}^\prime}{d^{\eta} } \right)
    \end{eqnarray*}
    or, equivalently,
    \begin{eqnarray*}
      \left(  d_{opt} \
        \Gamma\left( \frac{\bar{P_t}^\prime}{ d_{opt}^{\eta}} \right) \right)<
      x^{-\frac{1}{\eta}} d \ \Gamma\left(\frac{\bar{P_t}^\prime}{(x^{-\frac{1}{\eta}}
          d)^{\eta} } \right)
    \end{eqnarray*}
    which contradicts the hypothesis that $d_{opt}$ is optimal for
    $\bar{P_t}^\prime$.
    \item With the path loss model
	$\frac{P}{d^{\eta}}$, we see that for $d < d_0$, the received
	power is scaled more than the transmitted power $P$, due to the factor $\frac{1}{d^{\eta}}$,
	and an $d_0^{\eta}$ factor in $\alpha$, i.e., the model over-estimates
	the received power and the transport capacity. Hence, the achievable
	transport capacity for $d < d_0$
	is definitely less than $d \times \Gamma\left(\frac{\bar{P_t}^{\prime}}{d^{\eta}}\right)$.
	The result now follows from the scaling result in 1).

    \item It follows from 1) that, if $\bar{P_t}^\prime$ scales by a
      factor $x$, then the optimum $d$ scales by $x^{\frac{1}{\eta}}$,
      so that, at the optimum, $\frac{\bar{P_t}^{\prime}}{d^{\eta}}$ is
      unchanged. Hence the optimal $\{\xi(h)\}$ is unchanged, which means that
      $\{P(h)\}$ must scale by $x$. The water pouring form is evident.

    \item Again, by 1) and 2), if $\bar{P_t}^\prime$ scales by a factor $x$, then the
      optimum $d$ scales by $x^{\frac{1}{\eta}}$, so that, at the
      optimum, $\frac{\bar{P_t}^{\prime}}{d^{\eta}}$ is unchanged. Thus
      $\Gamma\left(\frac{\bar{P_t}^{\prime}}{d^{\eta}}\right) $ is
      unchanged, and the optimal transport capacity scales as the optimum $d$,
      i.e., by the factor $x^{\frac{1}{\eta}}$.
  \end{enumerate}

\end{proof}

\begin{remarks}
\end{remarks}
The above theorem yields the following observations for the fixed
transmission time model.
\begin{enumerate}
\item As an illustration, with $\eta = 3$, in order to double the
  transport capacity, we need to use $2^3$ times the
  $\bar{P_t}^{\prime}$. This would result in a considerable
  reduction in network lifetime, assuming the same battery energy.

\item We observe that as the power constraint $\bar{P_t}^\prime$ scales,
  the optimal bit rate carried in the network,
  $\Gamma\left(\frac{\bar{P_t}^{\prime}}{d^{\eta}}\right) $, stays
  constant, but the optimal transport capacity increases since the
  optimal hop length increases.  Further, because of the way the
  optimal power control and the optimal hop length scale together, the
  nodes transmit at the same \emph{physical bit rate} in each fading
  state; see the proof of Theorem~\ref{thm:fading_time} part 3).
\hfill \QED

\end{enumerate}

\subsection{Characterisation of the Optimal $d$}
\label{sec:characterise_d}
By the results in Theorem~\ref{thm:dgamma_properties} we can conclude
that the optimal solution of the maximisation in
(\ref{eqn:opt_over_d}) lies in the set of points for which the
derivative of $d \times
\Gamma\left(\frac{\bar{P_t}^{\prime}}{d^{\eta}}\right)$ is zero.  For a fixed $\bar{P_t}^{\prime}$,
define
$\pi(d) := \frac{\bar{P_t}^{\prime}}{d^{\eta}}$.  Differentiating $d
\times \Gamma(\pi(d))$, we obtain, (see Appendix~\ref{sec:stationary_points})
\begin{eqnarray*}
\frac{\partial}{\partial d}(d \ \Gamma(\pi(d)) &=&
\Gamma(\pi(d)) - \eta \pi(d) \lambda(\pi(d))
\end{eqnarray*}
where $\lambda(\pi)$ is the Lagrange multiplier for the optimisation
problem that yields $\Gamma(\pi(d))$.  Since $d$ appears
only via $\pi(d)$, we can view the right hand side as a function only of
$\pi$.  We are interested in the zeros of the above
expression. Clearly, $\pi = 0$ is a solution. The solution $\pi = 0$
corresponds to the case $d = \infty$;
However, we are interested only in solutions of $d$ in $(0, \infty)$, and hence,
we seek positive solutions of $\pi$ of
\begin{eqnarray}
\label{eqn:derivative_in_pi}
\Gamma(\pi) - \eta \pi \lambda(\pi) = 0
\end{eqnarray}

\begin{figure}[tb]
  \centering \
  \psfig{figure=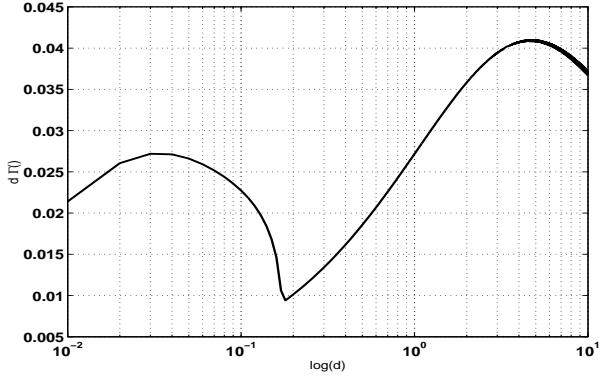,height=50mm,width=80mm}
  \caption{Plot of $d \times \Gamma\left(\frac{1}{d^3}\right)$ (linear
    scale) vs.\ $d$ (log scale) for a channel with two fading states
    $h_1, h_2$. The fading gains are $h_1 = 100$ and $h_2 = 0.5$, with
    probabilities $a_{h_1} = 0.01 = 1 - a_{h_2}$. The
    function has 3 non-trivial stationary points.}
  \label{fig:multimodal_example}
\end{figure}

\remarks{In Appendix~\ref{sec:stationary_points}, we consider a continuously
  distributed fading random variable $H$ with p.d.f. $a(h)$. The analysis can be done for
  a discrete valued fading distribution as well, and we provide this analysis in
  Appendix~\ref{sec:discrete_fading}.  The following example then
  illustrates that, in general, the function $\Gamma(\pi) - \eta \pi
  \lambda(\pi) = 0$ can have multiple solutions. Consider a fading
  distribution that takes two values: $h_1 = 100$ and $h_2 = 0.5$,
  with probabilities $a_{h_1} = 0.01 = 1 - a_{h_2}$.
  Figure~\ref{fig:multimodal_example} plots $d \times
  \Gamma\left(\frac{1}{d^{3}}\right)$ for the system with $\eta =
  3$. Notice that there are 3 stationary points other than the
  trivial solution $d = \infty$ (which is not shown in the figure).
  Also, the maximising solution is not
  the first stationary point (the stationary point close to $0$). If,
  on the other hand, $a_{h_1} = 0.001 = 1 - a_{h_2}$, we again have
  $3$ stationary points, but the optimal solution now is the first
  stationary point. \hfill \QED}

More generally, and still pursuing the discrete case, let ${\cal H}$
denote the set of fading states when the fading random variable is
discrete with a finite number of values; $|{\cal H}|$ denotes the
cardinality of ${\cal H}$.
\begin{theorem}
There are at most $2 |{\cal H}| - 1$ stationary points of
  $d \ \Gamma(\pi(d))$ in $0 < d < \infty$.
\end{theorem}
\begin{proof}
  See Appendix~\ref{sec:discrete_fading} for the related analysis and
  the proof of this theorem.
\end{proof}

We conclude from the above discussion that it is difficult to
characterise the optimal solution when there are multiple stationary
points. Hence we seek conditions for a unique positive stationary point,
which must then be the maximising solution. In
Appendix~\ref{sec:stationary_points},
 we have shown that the equation
characterising the stationary points, $\Gamma(\pi) - \eta \pi
\lambda(\pi) = 0$, can be rewritten as
\begin{eqnarray}
\label{eqn:stationary_point_rechar}
\int_{0}^{1} (\log(y) - \eta (y - 1)) \frac{\lambda^2}{y^2} f\left(\frac{\lambda}{y}\right) dy = 0
\end{eqnarray}
for $f(x) := a\left(\frac{\sigma^2 x}{\alpha}\right) \frac{\sigma^2}{\alpha}$, the density of
the random variable $X := \frac{\alpha H}{\sigma^2}$.  Notice that $\pi$
does not appear in this expression. The solution directly yields the Lagarange
multiplier of the throughput maximisation problem for the optimal
value of hop length.
The following theorem guarantees the existence of at most one solution of
(\ref{eqn:stationary_point_rechar}).
\begin{theorem}
\label{thm:unique_stationary_point}
If for
  any $\lambda_1 > \lambda_2 > 0$,
  $\frac{f\left(\frac{\lambda_2}{y}\right)}{f\left(\frac{\lambda_1}{y}\right)}$
  is a strictly monotonic decreasing function of $y$, then
  the objective function $d \times
  \Gamma\left(\frac{\bar{P_t}^{\prime}}{d^{\eta}}\right)$ has at most one
  stationary point $d_{opt}, 0 < d_{opt} < \infty$.
\end{theorem}
\begin{proof}
The proof follows from Lemmas~\ref{lem:stationary_points_lambda},
and \ref{lem:monotonicity_condition} in Appendix~\ref{sec:stationary_points}.
\end{proof}

\begin{corollary}
  If $H$ has an exponential distribution and $\eta \geq 2$, then
  the objective in the
  optimisation problem of (\ref{eqn:opt_over_d}) has a unique
  stationary point $d_{opt} \in (0,\infty)$, which achieves the maximum.
\end{corollary}
\begin{proof}
 $a(h)$ is of the form $\mu e^{- \mu h}$.
  From Theorem~\ref{thm:dgamma_properties},
  we see that $\lim_{d \to 0} d \times \Gamma\left(\frac{\bar{P_t}^{\prime}}{d^{\eta}}\right) = 0$ and $\lim_{d \to \infty} d \times \Gamma\left(\frac{\bar{P_t}^{\prime}}{d^{\eta}}\right) = 0$.
 And, the monotonicity hypothesis in
  Theorem~\ref{thm:unique_stationary_point} holds for $a(h)$.
\end{proof}

\begin{figure}[tb]
  \centering \
  \psfig{figure=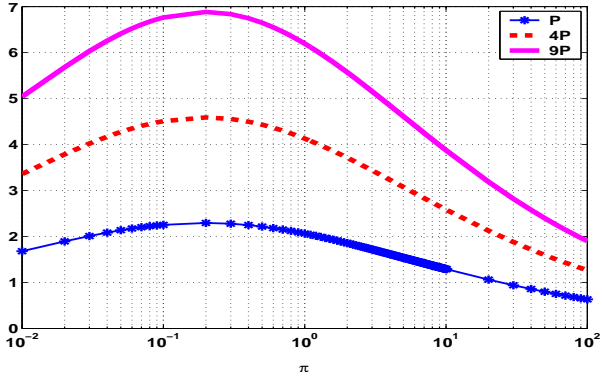,height=50mm,width=80mm}
  \caption{Plot of $d \times \Gamma\left(\frac{\bar{P_t}^{\prime}}{d^{\eta}}\right)$ (linear
	scale) vs. $\pi$ ($= \frac{\bar{P_t}^{\prime}}{d^{\eta}}$) (log scale) for a fading channel
	(with exponential distribution). We consider $3$ power levels ($\bar{P_t}^{\prime},
	4 \bar{P_t}^{\prime}$ and $9 \bar{P_t}^{\prime}$) and $\eta = 2$. The function has
	a unique optimum $\pi_{opt} (\pi_{opt} \approx 0.2)$ for all the $3$ cases.
         }
  \label{fig:exponential_example}
\end{figure}

\remarks{
  \begin{enumerate}
  \item Hence, for $\eta \geq 2$,
    for the Rayleigh fading model there exists a unique
    stationary point which corresponds to the optimal operating point.
 \item For $\bar{P_t}^{\prime} > \bar{P_t}^{\prime}_{\min}$, and for the conditions
  in Theorem~\ref{thm:dgamma_properties} and
  \ref{thm:unique_stationary_point}, let $\pi_{opt}$ denote the unique
  stationary point of (\ref{eqn:derivative_in_pi}). Then define $\Gamma(\pi_{opt}) = \Theta_{opt}$. It
  follows from Theorem~\ref{thm:fading_time} that the optimal transport capacity takes the
  form $\left(\frac{\bar{P_t}^{\prime}}{\pi_{opt}}\right)^{\frac{1}{\eta}} \Theta_{opt}$,
where $\Theta_{opt}$ depends on $a(h)$ and the MAC parameters but not on $\bar{P}$ (or $\bar{P_t}$).
\item Figure~\ref{fig:exponential_example} numerically illustrates our results for the Rayleigh
  fading distribution and $\eta = 2$. Scaling $\bar{P_t}^{\prime}$ by $4$ scales the transport
  capacity from $2.3$ to $4.6$, i.e., by $4^{\frac{1}{\eta}} = \sqrt{4}$ and
similarly for scaling $\bar{P_t}^{\prime}$ by $9$.

  \end{enumerate}
}
The uniqueness result guarantees that a distributed implementation of
the optimization problem, if it converges, shall converge to the
unique stationary point, which is the optimal solution.

\section{Conclusion}
\label{sec:conclusion}
In this paper we have studied a problem of optimal power
control and self-organisation in a single cell, dense, ad hoc multihop
wireless network.  The self-organisation is in terms of the
hop distance used when relaying packets between source-destination
pairs.

We formulated the problem as one of maximising the transport capacity
of the network subject to an average power constraint.  We showed
that, for a fixed transmission time scheme, there corresponds an
intrinsic aggregate packet carrying capacity at which the network
operates at the optimal operating point, independent of the average
power constraint. We also obtained the scaling law relating the
optimal hop distance to the power constraint, and hence relating the optimal
transport capacity to the power constraint (see
Theorem~\ref{thm:fading_time}). Because of the way the power control
and the optimal hop length scale, the optimal physical bit rate in
each fading state is invariant with the power constraint.
In Theorem~\ref{thm:unique_stationary_point} we provide a
characterisation of the optimal hop distance for cases in which the
fading density satisfies a certain monotonicity condition.

One motivation for our work is the optimal operation of sensor
networks. If a sensor network is supplied with external power, or if
the network is not required to have a long life-time, then the value
of the power constraint, $\bar{P}$, can be large, and a long hop distance will be used,
yielding a large transport capacity. On the other hand, if the sensor
network runs on batteries and needs to have a long life-time then
$\bar{P}$ would be small, yielding a small hop
length.  In either case, the optimal aggregate bit rate carried by the network
would be the same.

In \cite{srivathsa07self-organisation}, the author studies the problem
of developing a distributed algorithm for nodes to adapt themselves towards
the optimal operating point.
They first propose a distance discretization technique in which the hop distance on
the critical geometric graph is used as a distance measure.
Using the distance approximation, the author then develops a distributed algorithm
aimed to maximize the transport capacity of the network in the sense of the
framework presented in this paper.



\appendix
\renewcommand{\thelemma}{\thesubsection.\arabic{lemma}}
\renewcommand{\thetheorem}{\thesubsection.\arabic{theorem}}
\renewcommand{\theremarks}{\thesubsection.\arabic{remarks}}

\subsection{Stationary Points of $d \times \Gamma(\pi(d))$}
\label{sec:stationary_points}
Recall that we defined $\pi(d) := \frac{\bar{P_t}^\prime}{d^{\eta}}$.
Further, $\Gamma(\pi(d))$ was defined as
\begin{equation}
\label{eqn:define_Gamma}
\Gamma(\pi(d)) := \max \int_0^\infty \log\left(1 + \frac{\alpha h}{\sigma^2} \frac{P(h)}{d^\eta}\right) a(h) dh
\end{equation}
where the maximum is over all power controls $\{P(h)\}$ satisfying the
constraint
\begin{equation}
\label{eqn:pi_criterion}
\int_0^\infty \frac{P(h)}{d^\eta} a(h) dh \leq \pi(d)
\end{equation}
For ease of notation, let us use the substitution $x := \frac{\alpha
  h}{\sigma^2}$.  Write $\xi(x) := \xi(\frac{\alpha h}{\sigma^2}) = \frac{P(h)}{d^{\eta}}$
  and $f(x) := a\left(\frac{\sigma^2
    x}{\alpha}\right) \frac{\sigma^2}{\alpha}$.  Note that $f(\cdot)$
is the probability density of the random variable $X := \frac{\alpha
  H}{\sigma^2}$. Then, equations~(\ref{eqn:define_Gamma}) and
(\ref{eqn:pi_criterion}) can be  rewritten as
\[ \Gamma(\pi) = \max \int_0^\infty \log(1 + x \xi(x)) f(x) dx \]
and
\[ \int_0^\infty \xi(x) f(x) dx \leq \pi \]
This optimisation problem
is one of maximising a convex functional of $\{\xi(x)\}$, subject to
a linear constraint.  The optimal solution of the problem has
water-pouring form, and the optimal solution is given by,
\[ \xi(x) = \left(\frac{1}{\lambda(\pi)} - \frac{1}{x}\right)^+ \]
where $\lambda(\pi)$ is obtained from
\[ \int_{\lambda(\pi)}^{\infty} \left(\frac{1}{\lambda(\pi)} -
  \frac{1}{x}\right) f(x) dx = \pi \] Further, the derivative of the
optimum value $\Gamma(\pi)$, w.r.t.\ $\pi$, i.e., $\frac{\partial
  \Gamma(\pi)}{\partial \pi} = \lambda(\pi)$ (see Aubin~\cite{aubin00functional-analysis}).

Let us now reintroduce the dependence on $d$, and consider the problem of
maximizing $d \times \Gamma(\pi(d))$ over $d$.  Differentiating $d \times
\Gamma(\pi(d))$ w.r.t.\ $d$, we get,
\begin{eqnarray*}
  \frac{\partial}{\partial d} \left(d \ \Gamma(\pi(d))  \right)
  &=& \Gamma(\pi(d)) + d \ \frac{\partial}{\partial d} \Gamma(\pi(d)) \\
  &=& \Gamma(\pi(d)) + d \ \frac{\partial \Gamma}{\partial \pi} (\pi(d)) \times \frac{\partial \pi(d)}{\partial d} \\
  &=& \Gamma(\pi(d)) + d \ \Gamma^{\prime}(\pi(d)) \times \frac{- \eta \bar{P_t}^\prime}{d^{\eta + 1}} \\
&=& \Gamma(\pi(d)) - \eta \pi(d) \Gamma^{\prime}(\pi(d))
\end{eqnarray*}
where $\Gamma^\prime(\pi) := \frac{\partial \Gamma(\pi)}{\partial \pi}$.
 Substituting $\Gamma^{\prime}(\pi) = \lambda(\pi)$, we
have,
\begin{equation}
\label{eqn:stationary_points}
\frac{\partial}{\partial d}(d \Gamma(\pi(d))) =  \Gamma(\pi(d)) - \eta \pi(d) \lambda(\pi(d))
\end{equation}
The stationary points of $d \times \Gamma(\pi(d))$ are now obtained by
equating the right hand side of (\ref{eqn:stationary_points})
to zero.  Note that since $d$ appears in this equation only as
$\pi(d)$, we need only study the roots of the equation
\begin{eqnarray}
  \label{eqn:stationary_points_pi}
  \Gamma(\pi) - \eta \pi \lambda(\pi) = 0
\end{eqnarray}

We now proceed to obtain a characterisation of the stationary
points.  Substituting the optimal solution in the expression of
$\Gamma(\pi)$ and $\lambda(\pi)$, and suppressing the argument $\pi$
in $\lambda(\pi)$, we get,
\begin{equation}
\label{eqn:optimal_gamma}
\Gamma(\pi) = \int_{\lambda}^{\infty} \log\left(\frac{x}{\lambda}\right) f(x) dx
\end{equation}
with $\lambda$ being given by
\begin{equation}
\label{eqn:lambda_pi_relation}
\pi = \int_{\lambda}^{\infty} \left(\frac{1}{\lambda} -
  \frac{1}{x}\right) f(x) dx
\end{equation}
Using the substitution $z =
\frac{1}{x}$, $l = \frac{1}{\lambda}$, and defining $g(z) =
\frac{1}{z^2} f \left(\frac{1}{z}\right)$,
(\ref{eqn:optimal_gamma}) and (\ref{eqn:lambda_pi_relation}) becomes,
\begin{equation}
\label{eqn:gamma_l_z}
\Gamma(\pi) = \int_{0}^{l} \log\left(\frac{l}{z}\right) g(z) dz
\end{equation}
with $l$  (actually, $l(\pi)$) being given by
\begin{equation}
\label{eqn:pi_l_z}
\pi = \int_{0}^l \left(l - z\right) g(z) dz
\end{equation}
We note that $g(\cdot)$ is the density of the random variable $Z :=
\frac{1}{X} = \frac{\sigma^2}{\alpha H}$.

We will use the following definitions for convenience.
For a function $t(\cdot)$ of the random variable $Z$, define the operators
$\mathsf{E}_l(\cdot)$ and $\mathsf{G}_l(\cdot)$ as
\begin{eqnarray*}
  \mathsf{E}_l(t(Z)) &:=& \frac{\int_{0}^{l} t(z) g(z) dz}{\int_{0}^{l} g(z) dz}\\
  \mathsf{G}_l(t(Z)) &:=& \int_{0}^{l} t(z) g(z) dz
\end{eqnarray*}

\begin{lemma}
\label{lem:stationary_points_lambda}
  The roots of (\ref{eqn:stationary_points_pi}) are  equivalent
to the roots of the equation
\begin{eqnarray}
  \label{eqn:stationary_points_G}
 \eta \mathsf{G}_{l} \left(\frac{Z}{l} - 1\right) = \mathsf{G}_{l} \left( \log \left(\frac{Z}{l}\right) \right)
\end{eqnarray}
with $l$ then being given by (\ref{eqn:pi_l_z}).
\end{lemma}
\begin{proof}
Using the definitions of $\mathsf{E}_l(\cdot)$ and $\mathsf{G}_l(\cdot)$,
(\ref{eqn:gamma_l_z}) and (\ref{eqn:pi_l_z}) simplify to
\begin{equation}
\label{eqn:Gamma_simplified}
\Gamma(\pi) = \log(l) \mathsf{P}(Z\leq l) - \mathsf{G}_l(\log(Z))
\end{equation}
\begin{equation}
\label{eqn:pi_simplified}
\pi = l \mathsf{P}(Z \leq l) - \mathsf{G}_l(Z)
\end{equation}
(\ref{eqn:pi_simplified}) provides the $l$ (actually $l(\pi)$) to be substituted in
(\ref{eqn:Gamma_simplified}).  Substituting for $\Gamma(\pi)$
(from (\ref{eqn:Gamma_simplified})), and for $l$ (from
(\ref{eqn:pi_simplified})), into
(\ref{eqn:stationary_points_pi}), dividing across by $\mathsf{P}(Z \leq
l)$, and using the definition of $\mathsf{E}_l(\cdot)$, we have,
\[ \log\left(\frac{\pi + \mathsf{G}_l(Z)}{\mathsf{P}(Z \leq l)}\right) - \mathsf{E}_l(\log(Z)) - \frac{\eta \pi}{\pi + \mathsf{G}_l(Z)} = 0 \]
\[ \log\left(\frac{\pi}{\mathsf{P}(Z \leq l)} + \mathsf{E}_l(Z)\right) - \mathsf{E}_l(\log(Z)) - \frac{\eta \pi}{\pi + \mathsf{G}_l(Z)} = 0 \]
\begin{eqnarray*}
\log\left[\left(\frac{\pi}{\mathsf{G}_l(Z)} + 1\right) \mathsf{E}_l(Z)\right] + \log\left(e^{- \mathsf{E}_l(\log(Z))}\right)  \\
- \frac{\eta \pi}{\pi + \mathsf{G}_l(Z)} = 0
\end{eqnarray*}
Rearranging terms, we get,
\begin{eqnarray*}
\log\left(\frac{\pi + \mathsf{G}_l(Z)}{\mathsf{G}_l(Z)}\right) +
\log\left(\mathsf{E}_l(Z) e^{- \mathsf{E}_l(\log(Z))}\right) \\
- \frac{\eta \pi}{\pi + \mathsf{G}_l(Z)} = 0
\end{eqnarray*}
Denote $b_l :=
\log\left(\mathsf{E}_l(Z) e^{- \mathsf{E}_l(\log(Z))}\right)$.  Then,
we have,
\[ \log\left(\frac{\pi + \mathsf{G}_l(Z)}{\mathsf{G}_l(Z)}\right) + b_l  - \frac{\eta \pi}{\pi + \mathsf{G}_l(Z)} = 0 \]
From (\ref{eqn:pi_simplified}), we have
\[ \frac{\mathsf{G}_l(Z)}{\pi + \mathsf{G}_l(Z)} = \frac{\mathsf{G}_l(Z)}{l \mathsf{P}(Z \leq l)} = \frac{\mathsf{E}_l(Z)}{l} \]
which, with the previous equation, yields
\[ \log\left(\frac{l}{\mathsf{E}_l(Z)}\right) + b_l - \eta \left( 1 -
  \frac{\mathsf{E}_l(Z)}{l}\right) = 0 \] Recall that $l$ is actually
$l(\pi)$. We now find that $\pi$ appears in the equation only as
$l(\pi)$. Hence we can view this as an equation in the variable $l (=
\frac{1}{\lambda})$.  Rearranging terms, we get
\[ - \log\left(\frac{\mathsf{E}_l(Z)}{l}\right) + \eta
\frac{\mathsf{E}_l(Z)}{l} = - (b_l - \eta) \]
Exponentiating both
sides, and substituting back for $b_l$, yields
\[ \frac{\mathsf{E}_l(Z)}{l} e^{- \eta \frac{\mathsf{E}_l(Z)}{l}} = \mathsf{E}_l(Z) e^{- \mathsf{E}_l(\log(Z))} e^{-\eta} \]
On cancelling $\mathsf{E}_l(Z)$, and transposing terms, we next obtain
\[ e^{- \eta \left( \frac{\mathsf{E}_l(Z)}{l} - 1 \right)} = e^{- \mathsf{E}_l\left(\log\left(\frac{Z}{l}\right)\right)} \]
or,
\[ e^{- \eta \left(\mathsf{E}_l\left(\frac{Z - l}{l}\right)\right)} = e^{- \mathsf{E}_l\left(\log\left(\frac{Z}{l}\right)\right)} \]
Taking $\log$ on both sides, we have,
\[ \eta \mathsf{E}_l\left(\frac{Z-l}{l}\right) =
\mathsf{E}_l\left(\log\left(\frac{Z}{l}\right)\right) \] In terms of
$\mathsf{G}_l(\cdot)$, this is equivalent to
\[ \eta \mathsf{G}_l\left(\frac{Z-l}{l}\right) =
\mathsf{G}_l\left(\log\left(\frac{Z}{l}\right)\right)\] which is the
desired result.
\end{proof}

We next address the question of a \emph{unique} positive solution of
(\ref{eqn:stationary_points_G}). The following lemma
guarantees the existence of a unique positive solution, when
$f(\cdot)$, the density of $\frac{\alpha H}{\sigma^2}$, satisfies a
certain monotonicity condition.
\begin{lemma}
\label{lem:monotonicity_condition}
   (\ref{eqn:stationary_points_G}) has at most one
  positive solution if for any $0 < l_1 < l_2$,
  $\frac{f\left(\frac{1}{y l_2}\right)}{f\left(\frac{1}{y l_1}\right)}$
  is a strictly monotone decreasing function of $y$.
\end{lemma}
\begin{proof}
 Expanding
$\mathsf{G}_{l}(\cdot)$, (\ref{eqn:stationary_points_G}) becomes,
\[\eta \int_{0}^{l} \left( \frac{z}{l} - 1 \right) g(z) dz -
\int_{0}^{\frac{1}{\lambda}} \log\left(\frac{z}{l}\right) g(z) dz =0\]
Rewriting the equation in terms of $f(\cdot)$, we have,
\begin{eqnarray*}
\int_{0}^{l} \left( \eta \left( \frac{z}{l} - 1 \right) - \log\left(\frac{z}{l}\right) \right) \frac{1}{z^2}
f\left(\frac{1}{z}\right) dz = 0
\end{eqnarray*}
Using a substitution $y = \frac{z}{l}$ in the above equation, we get,
\begin{eqnarray}
  \label{eqn:stationary_points_y}
\int_{0}^{1} (\log(y) - \eta (y - 1)) \frac{1}{y^2 l^2} f\left(\frac{1}{y l}\right) dy = 0
\end{eqnarray}
Define $c(y) := (\log(y)  - \eta (y - 1)) \frac{1}{y^2}$ and
$b_{l}(y) := f\left(\frac{1}{y l}\right)$.
We are now interested in a positive $l$ that solves
\[ \int_{0}^{1} c(y) b_{l}(y) dy = 0 \]
Observe that $\lim_{y \to 0} c(y) = -\infty$ and $c(1) = 0$. Further,
there exists a unique $y^{\prime}$ such that $c(y) \leq 0$ for all $0
\leq y \leq y^{\prime}$ and $c(y) \geq 0$ for all $y^{\prime} \leq y
\leq 1$.  Since $b_{l}(y) \geq 0$ for all $y$ and $l$, we
have $c(y) b_{l}(y) \leq 0$ for all $0 \leq y \leq y^{\prime}$
and $c(y) b_{l}(y) \geq 0$ for all $y^{\prime} \leq y \leq 1$. In particular,
\[ \int_{0}^{y^{\prime}} c(y) b_l(y) dy \leq 0 \]
\[ \int_{y^{\prime}}^1 c(y) b_l(y) dy \geq 0 \]

Consider $l_1, l_2$ such that $0 < l_1 < l_2$.
By hypothesis, $\frac{b_{l_2}(y)}{b_{l_1}(y)}$ is a strictly monotone
decreasing function of $y$.
Hence, $\frac{c(y) b_{l_2}(y)}{c(y) b_{l_1}(y)}$ is
also a strictly monotone decreasing function of $y$. We then have,
\[ \frac{\int_{0}^{y^{\prime}} |c(y)| b_{l_2}(y) dy}{\int_{0}^{y^{\prime}} |c(y)| b_{l_1}(y) dy}
   = \frac{\int_{0}^{y^{\prime}} |c(y)| \frac{b_{l_2}(y)}{b_{l_1}(y)} b_{l_1}(y) dy}{\int_{0}^{y^{\prime}} |c(y)|
b_{l_1}(y) dy}
 > \frac{b_{l_2}(y^{\prime})}{b_{l_1}(y^{\prime})},  \]
And,
\[ \frac{\int_{y^{\prime}}^{1} c(y) b_{l_2}(y) dy}{\int_{y^{\prime}}^{1} c(y) b_{l_1}(y) dy} =
        \frac{\int_{y^{\prime}}^{1} c(y) \frac{b_{l_2}(y)}{b_{l_1}(y)} b_{l_1}(y) dy}{\int_{y^{\prime}}^{1} c(
y) b_{l_1}(y) dy}
< \frac{b_{l_2}(y^{\prime})}{b_{l_1}(y^{\prime})} \]
Hence,
\[ \frac{\int_{0}^{y^{\prime}} |c(y)| b_{l_2}(y) dy}{\int_{0}^{y^{\prime}} |c(y)| b_{l_1}(y) dy}
> \frac{\int_{y^{\prime}}^{1} c(y) b_{l_2}(y) dy}{\int_{y^{\prime}}^{1} c(y) b_{l_1}(y) dy}\]
Interchanging terms, we get,
\[ \frac{\int_{0}^{y^{\prime}} |c(y)| b_{l_2}(y) dy}{\int_{y^{\prime}}^{1} c(y) b_{l_2}(y) dy} >
   \frac{\int_{0}^{y^{\prime}} |c(y)| b_{l_1}(y) dy}{\int_{y^{\prime}}^{1} c(y) b_{l_1}(y) dy} \]
i.e., the ratio of the negative area of the integral to the positive area of the integral is a strictly monotonic function
of $l$.
Hence, as $l$ increases, the integral (\ref{eqn:stationary_points_y}) can cross $0$ at most once,
or, there exists at most one (non-trivial) solution for (\ref{eqn:stationary_points_y}).
\end{proof}

\subsection{Proof of Theorem~\ref{thm:dgamma_properties}}
\label{sec:dgamma_limits}
In this section, we will use the variables and equations from
the discussion in Appendix~\ref{sec:stationary_points}.

\begin{lemma}
\label{lem:dgamma_continuous_diff}
$d \times \Gamma\left(\frac{\bar{P_t}^{\prime}}{d^{\eta}}\right)$ is continuously
differentiable with respect to $d$.
\end{lemma}
\proof{
Recall that $\pi := \frac{\bar{P_t}^{\prime}}{d^{\eta}}$.
$\Gamma\left(\pi\right)$ and $\lambda(\pi)$ (equations (\ref{eqn:optimal_gamma})
and (\ref{eqn:lambda_pi_relation})) are
continuous function of $\pi$, and $\pi$
itself is a continous function of $d$. Hence, from (\ref{eqn:stationary_points}),
we see that $d \times \Gamma\left(\frac{\bar{P_t}^{\prime}}{d^{\eta}}\right)$
is a continously differentiable function of $d$.
\hfill \QED}

\begin{lemma}
\label{lem:dgamma_zero}
If $H$ (or equivalently $X := \frac{H \alpha}{\sigma^2}$) has a finite mean,
then $\lim_{d \to 0} d \times \Gamma\left(\frac{\bar{P_t}^{\prime}}{d^{\eta}}\right) = 0$.
\end{lemma}
\begin{proof}
Consider (\ref{eqn:pi_l_z})
\[ \int_{0}^{l} (l - z) g(z) dz = \pi \]
where $l$ is in fact $l(\pi)$.
Talking $l$ outside the integral, we get,
\[ l \int_{0}^{l} \left(1 - \frac{z}{l} \right) g(z) dz = \pi\]
Rewriting the integral as an expectation,
we have, \mbox{$l \ \mathbf{E}_{z}\left( 1 - \frac{Z}{l}\right)^+ = \pi$}
or \mbox{$\mathbf{E}_{z}\left( 1 - \frac{Z}{l}\right)^+ = \frac{\pi}{l}$}.
Using Monotonce Convergence Theorem,
we get,
\[ \lim_{l \to \infty} \mathbf{E}_z\left(1 - \frac{Z}{l}\right)^+ \uparrow 1 \]
or,
\[ \lim_{l \to \infty} \frac{\pi}{l} = 1 \]
From (\ref{eqn:pi_l_z}), we see that, $l \to \infty$ as $\pi \to \infty$ ($d \to 0$).
Hence, we have,
\begin{equation}
\label{eqn:lim_l_pi}
\lim_{\pi \to \infty} \frac{l(\pi)}{\pi} = 1
\end{equation}
Now, consider the following limit, $\lim_{d \to 0} d \times \Gamma(\pi(d))$, or equivalently,
$\lim_{\pi \to \infty} \pi^{- \frac{1}{\eta}} \Gamma(\pi)$.
We know that,
\[ \pi^{- \frac{1}{\eta}} \Gamma(\pi) \geq 0 \]
From (\ref{eqn:gamma_l_z}), we have,
\[ \pi^{- \frac{1}{\eta}} \Gamma(\pi) = \pi^{- \frac{1}{\eta}} \mathbf{E}_z\left( - \log\left(\frac{Z}{l(\pi)}\right) \right)^+ \]
Expanding
the term inside the expectation, we have,
\[ = \pi^{- \frac{1}{\eta}} \mathbf{E}_z\left( \log\left(\frac{1}{Z}\right) + \log\left(\frac{l(\pi)}{\pi}\right) + \log(\pi) \right)^+ \]
Using the inequality $\log\left(\frac{1}{z}\right) \leq \frac{1}{z}$ (for $z \geq 0$)
in the above inequality, we get,
\[ \leq \pi^{- \frac{1}{\eta}} \mathbf{E}_z\left( \frac{1}{Z} + \log\left(\frac{l(\pi)}{\pi}\right) + \log(\pi)\right)^+ \]
$\mathbf{E}_z\left(\frac{1}{Z}\right) < \infty$
(follows from the definition $Z := \frac{1}{X}$ and the hypothesis on $\mathbf{E}X$), $\eta > 0$
and from (\ref{eqn:lim_l_pi}), we have the right hand side of the above
expression $\to 0$ as $\pi \to \infty$, which implies that $\lim_{\pi \to \infty} \pi^{- \frac{1}{\eta}} \Gamma(\pi) = 0$, or
\[ \lim_{d \to 0} d \times \Gamma(\pi(d)) = 0 \]
\end{proof}

\begin{lemma}
\label{lem:dgamma_infinity_slope}
Let $\eta \geq 2$, $\frac{1}{x^2} f\left(\frac{1}{x}\right)$ be continuously
differentiable and $\lim_{x \to 0} \frac{1}{x^2} f\left(\frac{1}{x}\right) = 0$. Then
$\frac{\partial}{\partial d} \left(d \times \Gamma\left(\frac{\bar{P_t}^{\prime}}{d^{\eta}}\right)\right)
\leq 0$ as $d \to \infty$.
\end{lemma}
\begin{proof}
From (\ref{eqn:stationary_points}) and the discussion in the
proof of Lemma~\ref{lem:stationary_points_lambda}, we have,
\begin{eqnarray*}
\frac{\partial}{\partial d}(d \ \Gamma(\pi(d))) =  \Gamma(\pi(d)) - \eta \pi(d) \lambda(\pi(d)) \\
= \kappa \int_{0}^{l} \left(\eta \left( \frac{z}{l} - 1\right) - \log\left( \frac{z}{l} \right) \right) \frac{1}{z^2} f\left(\frac{1}{z}\right) dz \\
\end{eqnarray*}
where $\kappa \geq 0$. Using a substitution $y = \frac{z}{l}$, we get,
$\frac{\partial}{\partial d}(d \ \Gamma(\pi(d)))$
\begin{eqnarray}
\label{eqn:derivative_in_yl}
= \kappa \int_{0}^{1} (\eta (y - 1) - \log(y)) \frac{1}{y^2 l^2} f\left(\frac{1}{y l}\right) dy
\end{eqnarray}
Define $b(y) := \eta (y - 1) - \log(y)$.
For $\eta > 1$, there exists a $y^{\prime}$ (depending on $\eta$)
such that $b(y) \geq 0$ for $0 \leq y \leq y^{\prime}$ and
$b(y) \leq 0$ for $y^{\prime} \leq y \leq 1$, also
$b(1) = 0$.
Then, in (\ref{eqn:derivative_in_yl}), we see that,
\begin{eqnarray*}
\int_{0}^{y^{\prime}} (\eta (y - 1) - \log(y)) \frac{1}{y^2} f\left(\frac{1}{y l}\right) dy \geq 0 \\
\int_{y^{\prime}}^{1} (\eta (y - 1) - \log(y)) \frac{1}{y^2} f\left(\frac{1}{y l}\right) dy \leq 0
\end{eqnarray*}
Further,
\[ \int_{0}^{1} (\eta (y - 1) - \log(y)) dy = 1 - \frac{\eta}{2} \]
For $\eta \geq 2$, the integral $\int_{0}^1 b(y) dy$ is non-positive.

Let $g(y) := \frac{1}{y^2} f\left(\frac{1}{y}\right)$. Then $g(y)$ is  continuously
differentiable function and $\lim_{y \to 0} g(y) = 0$ (by hypothesis).
Define $y_0$ as
\[ y_0 := \sup\{y : g(z) = 0,  0 \leq z \leq y\} \]
If $y_0 > 0$, then, we see that for $l$ sufficiently small,
\[ \int_{0}^{y^{\prime}} (\eta (y - 1) - \log(y)) \frac{1}{y^2 l^2} f\left(\frac{1}{y l}\right) dy = 0 \]
This is because for sufficiently small $l$,
$\frac{1}{y^2} f\left(\frac{1}{y l}\right) = 0$ for $0 \leq y \leq y^{\prime}$.
Hence, $\lim_{d \to \infty} \frac{\partial}{\partial d} \left(d \times \Gamma\left(\frac{\bar{P_t}^{\prime}}{d^{\eta}}\right)\right) \leq 0$.

If $y_0 = 0$, we then have
$g^{\prime}(y) \geq 0$ in a small neighbourhood of $0$ (since $g$ is continuously differentiable
by hypothesis).
Hence, the function $g(y)$ is a monotonic increasing function in an $\epsilon$
neighbourhood of $0$,
i.e., $g(0) < g(y) \leq g(y^{\prime}) \leq g(\epsilon)$
for all $0 < y < y^{\prime} < \epsilon$.
Hence for all sufficiently small $l$, $\frac{1}{y^2} f(\frac{1}{y l})$
is a monotone increasing function of y in $[0, 1]$. Hence, in (\ref{eqn:derivative_in_yl}), we have,
\begin{eqnarray*}
\int_{0}^{y^{\prime}} (\eta (y-1) - \log(y)) \frac{1}{y^2 l^2} f\left(\frac{1}{y l}\right) dy +\\
 \int_{y^{\prime}}^{1} (\eta(y-1) - \log(y)) \frac{1}{y^2 l^2} f\left(\frac{1}{y l}\right) dy
\end{eqnarray*}
\[\leq  \left(\frac{1}{y^{\prime} l}\right)^2 f\left(\frac{1}{y^{\prime} l}\right) \int_{0}^{y^{\prime}} (\eta (y-1) - \log(y)) dy + \]
\[ +  \left(\frac{1}{y^{\prime} l}\right)^2 f\left(\frac{1}{y^{\prime} l}\right) \int_{y^{\prime}}^{1} (\eta (y-1) - \log(y)) dy \]
\[ = \left(\frac{1}{y^{\prime} l}\right)^2 f\left(\frac{1}{y^{\prime} l}\right) \left(1 - \frac{\eta}{2} \right)\]
The final expression is non-positive for $\eta \geq 2$.
Thus, $\frac{\partial}{\partial d} \left(d \times \Gamma\left(\frac{\bar{P_t}^{\prime}}{d^{\eta}}\right)\right) \leq 0$ as $d \to \infty$.
\end{proof}

\begin{lemma}
\label{lem:dgamma_infinity}
Let $\eta \geq 2$ and $\frac{1}{x^2} f\left(\frac{1}{x}\right)$ be continuously
differentiable. If for large $x$, $\mathsf{P}(X > x) = O(\frac{1}{x^2})$ (or equivalently
for $H = \frac{\sigma^2 X}{\alpha}$), then  $\lim_{d \to \infty}
d \times \Gamma\left(\frac{\bar{P_t}^{\prime}}{d^{\eta}}\right) = 0$.
\end{lemma}
\begin{proof}
Let $\mathsf{P}(X > x) = O(\frac{1}{x^2})$ for large $x$.
i.e.,
\[ \int_{x}^{\infty} f(x) dx = O\left(\frac{1}{x^2}\right) \]
Using a substitution $z = \frac{1}{x}$, we have,
\[ \int_{0}^{z} \frac{1}{z^2} f\left(\frac{1}{z}\right) dz = O(z^2)\]
Define $g(z) := \frac{1}{z^2} f\left(\frac{1}{z}\right)$.
Then,
\begin{equation}
\label{eqn:g_zero_zero}
\int_{0}^{z} g(z) dz = O(z^2)
\end{equation}
Since $g(z) \geq 0$ and continuous (by hypothesis), we have, $g(0) = 0$.
Suppose not, then, we
have $g(z) \geq \epsilon$ for all $0 \leq z < \delta$ for
some $\delta$.
Then,
\[ \int_{0}^{z} g(z) dz \geq \epsilon z \]
for all $z \leq \delta$, which is a contradiction to (\ref{eqn:g_zero_zero}).
Hence $\lim_{z \to 0} g(z) = 0$ or $\lim_{z \to 0} \frac{1}{z^2} f\left(\frac{1}{z}\right) = 0$.

\noindent
We know from (\ref{eqn:stationary_points}) that
\[ \frac{\partial}{\partial d} (d \ \Gamma(\pi(d))) = \Gamma(\pi(d)) - \eta \pi(d) \lambda(\pi(d)) \]
Now from Lemma~\ref{lem:dgamma_infinity_slope}, we see that, for $\eta \geq 2$, and for $d \to \infty$,
\begin{eqnarray*}
\Gamma(\pi(d)) - \eta \pi(d) \lambda(\pi(d)) &\leq& 0
\end{eqnarray*}
In other words,
\[ \Gamma(\pi(d)) \leq \eta \pi(d) \lambda(\pi(d)) \]
Multiplying by $d$ on both the sides, we have,
\begin{eqnarray}
\label{eqn:d_lambda_relation}
d \ \Gamma(\pi(d)) \leq \eta \pi(d) \lambda(\pi(d)) d = \eta \frac{\bar{P_t}^{\prime}}{d^{\eta - 1}} \lambda(\pi(d))
\end{eqnarray}
Since $\frac{\partial}{\partial d} \left(d \ \Gamma\left(\frac{\bar{P_t}^{\prime}}{d^{\eta}}\right)\right) \leq 0$ as $d \to \infty$,
the function $d \ \Gamma(\pi(d))$ is monotonic decreasing for $d \to \infty$.
Also $d \ \Gamma(\pi(d)) \geq 0$.
Suppose that, $\lim_{d \to \infty} d \ \Gamma(\pi(d)) \neq 0$, it implies that
$\lim_{d \to \infty} d \ \Gamma(\pi(d)) \geq \epsilon > 0$,
which, using (\ref{eqn:d_lambda_relation}), implies that $\frac{\lambda(\pi(d))}{d^{\eta - 1}} \geq \epsilon$ or as $d \to \infty$
\begin{eqnarray}
\label{eqn:lambda_d_infty}
\lambda(\pi(d)) \geq \epsilon d^{\eta - 1}
\end{eqnarray}

From (\ref{eqn:lambda_pi_relation}), we have,
\[ \int_{\lambda}^{\infty} \left( \frac{1}{\lambda} - \frac{1}{x} \right) f(x) dx = \frac{\bar{P_t}^{\prime}}{d^{\eta}} \]
ignoring the negative term, we have,
\begin{eqnarray*}
\frac{1}{\lambda} \int_{\lambda}^{\infty} f(x) dx &\geq& \frac{\bar{P_t}^{\prime}}{d^{\eta}}
\end{eqnarray*}
or,
\begin{eqnarray*}
\int_{\lambda}^{\infty} f(x) dx &\geq& \frac{\bar{P_t}^{\prime}}{d^{\eta}} \lambda
\end{eqnarray*}
Substituting from (\ref{eqn:lambda_d_infty}), we have,
\begin{eqnarray}
\int_{\lambda}^{\infty} f(x) dx &\geq& \frac{\bar{P_t}^{\prime}}{d^{\eta}} \epsilon d^{\eta - 1}
= \bar{P_t}^{\prime} \epsilon \frac{1}{d}
\label{eqn:tail_f_1}
\end{eqnarray}
But we have
\begin{equation}
\label{eqn:tail_f_2}
\int_{\lambda}^{\infty} f(x) dx = \mathsf{P}(X > \lambda) =  O\left(\frac{1}{\lambda^2}\right) \leq O\left(\frac{1}{d^{2 \eta - 2}}\right)
\end{equation}
where the last inequality follows from (\ref{eqn:lambda_d_infty}).
For $\eta \geq 2$, (\ref{eqn:tail_f_1}) and (\ref{eqn:tail_f_2}) yields a contradiction.
Hence, $\lim_{d \to \infty} d \times \Gamma\left(\frac{\bar{P_t}}{d^{\eta}}\right) = 0$.
\end{proof}

\newpage

\subsection{Discrete Fading States}
\label{sec:discrete_fading}
The optimization problem (\ref{eqn:optimization_sub})
for the discrete fading state case, simplifies to
\begin{eqnarray}
  \label{eqn:fixed_T_optimal_rate_problem}
 \max \  \ \sum_{h \in {\cal H}} a_h
    \ln\left(1 + \left(\frac{\alpha h}{\sigma^2}\right)
    \frac{P(h)}{d^\eta} \right) \nonumber\\
\mbox{subject to} \  \
 \sum_{h \in {\cal H}} a_h P(h) \leq \bar{P_t}^{\prime}
\end{eqnarray}
For notational convenience, let us index the set of fading states, ${\cal H}$,
in descending order by the index $i, 1 \leq i \leq |{\cal H}|$, i.e.,
$h_1 > h_2 > h_3 > \cdots$. Further,
denote
\[ a_{h_i} = a_i,\  x_i = \frac{\alpha h_i}{\sigma^2}, \mbox{and}\   \xi_i = \frac{P(h_i)}{d^\eta} \]
Also, denote
\[ \Pi = \frac{\bar{P_t}^{\prime}}{d^\eta} \]
We will later recall that, for each power constraint
$\bar{P_t}^{\prime}$,  $\Pi$ is a function of $d$.
Using this new notation and  change of variables, we obtain the problem
\begin{eqnarray}
  \label{eqn:fixed_T_optimal_rate_problem_newvar}
  \max \ \  \sum_{i} a_i \ln \left(1 + x_i \xi_i \right)  \nonumber \\
  \mbox{subject to} \ \  \sum_i a_i \xi_i \leq \Pi
\end{eqnarray}

We have the maximisation of a concave mapping from $\mathds{R}^{|{\cal H}|}$ to $\mathds{R}$ subject to a linear constraint. The KKT conditions are necessary
and sufficient, and the following ``water pouring'' form of the optimal solution is well
known. There exists $\lambda(\Pi) > 0$, such that,
for $ 1\leq i \leq |{\cal H}|$,
\[ \xi_i = \left(\frac{1}{\lambda(\Pi)} - \frac{1}{x_i} \right)^+ \]
with $\lambda(\Pi)$ being given by
\[ \sum_{\{i: \frac{x_i}{\lambda(\Pi)} > 1\}} a_i \left(\frac{1}{\lambda(\Pi)} - \frac{1}{x_i} \right) = \Pi \]
Defining,  for $1 \leq k \leq |{\cal H}|$,
\[ p_k = a_1 + a_2 + \cdots + a_k, \ \mbox{and}
                \ \alpha_k = \sum_{i=1}^k \frac{a_i}{x_i} \]
and $\Pi_0 = 0, \Pi_{|{\cal H}|} = \infty$, the Lagrange multiplier,
$\lambda(\Pi)$, is given by
\begin{eqnarray}
  \label{eqn:lambda_Pi}
\lambda(\Pi)=\left( \frac{1}{p_k} \left(\alpha_k + \Pi \right) \right)^{-1}
\end{eqnarray}
for $\Pi_{k-1} < \Pi \leq \Pi_{k}$ when $1 \leq k \leq |{\cal H}| - 1$,
and for $\Pi_{|{\cal H}| -1} < \Pi < \infty$ when $k = |{\cal H}|$.
Here the break-points  $\Pi_k, 1 \leq k \leq  |{\cal H}| - 1$, are obtained by
equating the values of $\lambda(\Pi)$  on either sides of the break-points,
and are expressed as
\[ \Pi_k = \left( \frac{\frac{\alpha_{k+1}}{p_{k+1}} - \frac{\alpha_k}{p_k}}{
                 \frac{1}{p_k} - \frac{1}{p_{k+1}}} \right) \]
The denominator of this expression is clearly $> 0$, and
a little algebra shows that, since $x_{k+1} > x_i, 1 \leq i \leq k$,
the numerator is also $> 0$.

For each $\Pi$, let us denote the optimal value of the problem defined
by (\ref{eqn:fixed_T_optimal_rate_problem_newvar}) by $\Gamma(\Pi)$.
We infer that
\[ \frac{\partial \Gamma}{\partial \Pi} = \lambda(\Pi) \]
Now, fixing the power constraint $\bar{P_t}^{\prime}$, and
 reintroducing the dependence on $d$,  we recall that
$\Pi(d) = \frac{\bar{P_t}^{\prime}}{d^\eta}$, and hence conclude that
\[ \frac{\partial \Gamma}{\partial d} = \lambda(\Pi(d))
           \left( \frac{-\eta \bar{P_t}^{\prime}}{d^{\eta+1}} \right) \]
Define $d_0 = \infty, d_{|{\cal H}|} = 0$, and, for $1 \leq k
\leq{|{\cal H}|} -1$, define
\[ d_k^{\eta} =\bar{P_t}^{\prime} \cdot \left(\frac{\frac{1}{p_k} - \frac{1}{p_{k+1}}}{
           \frac{\alpha_{k+1}}{p_{k+1}}-\frac{\alpha_k}{p_k}} \right) \]
Note that
$0 = d_{|{\cal H}|} < d_{|{\cal H}|-1} < \cdots < d_2 <  d_1 < d_0 = \infty$.
Now, substituting for
$\lambda(\Pi(d))$ from (\ref{eqn:lambda_Pi}) and integrating, yields
the following result
\begin{theorem}
\label{thm:characterisation_of_gamma_d}
For given $\bar{P_t}^{\prime}$, the optimal value $\Gamma(d)$ of the problem
defined by (\ref{eqn:fixed_T_optimal_rate_problem}) has the following
characterisation.
\begin{enumerate}
\item The derivative of $\Gamma(d)$ w.r.t.\ $d$ is given by
\begin{eqnarray}
  \label{eqn:derivative_of_optimal_rate_wrt_d}
  \frac{\partial \Gamma}{\partial d} &=& \frac{1}{d}
         \left( \frac{-\eta p_k \bar{P_t}^{\prime}}{\alpha_k d^\eta + \bar{P_t}^{\prime}} \right)
\end{eqnarray}
for $d_k \leq d < d_{k-1}$ when $1 \leq k \leq |{\cal H}| - 1$, and
for $0 < d < d_{|{\cal H}| -1}$ when $k = {|{\cal H}|}$.

\item $\frac{\partial \Gamma}{\partial d} $ is a negative, continuous
  and increasing function of $d$. In particular $\Gamma(d)$ is a
  decreasing, and convex function of $d$.

\item The function $\Gamma(d)$ is given by
\begin{eqnarray}
  \label{eqn:optimal_rate_as_fn_of_d}
  \Gamma(d) = p_k \ln\left(\alpha_k + \frac{\bar{P_t}^{\prime}}{d^\eta}\right) \gamma_k
\end{eqnarray}
for $d_k \leq d < d_{k-1}$ when $1 \leq k \leq |{\cal H}| - 1$, and
for $0 < d < d_{|{\cal H}| -1}$ when $k = {|{\cal H}|}$, with the
constants of integration $\gamma_k$ being given as follows.
\[ \gamma_1 = \frac{1}{\alpha_1} = \frac{x_1}{a_1} \]
and, for $2 \leq k \leq {\cal H}$, $\gamma_k$ is obtained recursively as
\[ \gamma_k = \frac{\left( \left(\alpha_{k-1} +
           \frac{\bar{P_t}^{\prime}}{d_{k-1}^\eta} \right) \gamma_{k-1} \right)^{
  \left(\frac{p_{k-1}}{p_k}\right)}}{\alpha_k +
       \frac{\bar{P_t}^{\prime}}{d_{k-1}^\eta}}  \]
\end{enumerate}
\end{theorem}
\proof{(\ref{eqn:optimal_rate_as_fn_of_d}) is obtained by
  integrating the derivative in
  (\ref{eqn:derivative_of_optimal_rate_wrt_d}) over each
  segment of its definition. The integration constants $\gamma_k$ are
  obtained by equating $\Gamma(d)$ on either sides of the break-points
  of the argument $d$.} \hfill \QED

\subsubsection{Optimisation over $d$}
\begin{figure}[tb]
  \centering \
  \psfig{figure=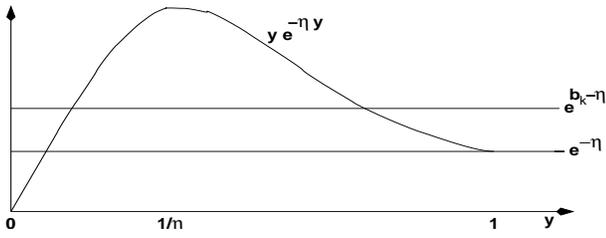,height=30mm,width=80mm}
  \caption{The stationary points of $\Gamma(d) d$ lie among the
    intersections of the curve $y e^{-\eta y}$ and lines $e^{b_k -
      \eta}, 1 \leq k \leq |{\cal H}|$, in the interval $0 < y < 1$.
    Here the plot is drawn for $\eta = 3$.}
  \label{fig:gamma_d_stationary_points}
\end{figure}

Using Theorem~\ref{thm:characterisation_of_gamma_d}, we conclude that
we need to look at the stationary points of $\Gamma(d) d$. To this
end, consider the solutions of
\[ \Gamma(d) + d \  \Gamma^{\prime} (d) = 0 \]
Reintroducing the variable $\Pi = \frac{\bar{P_t}^{\prime}}{d^\eta}$, and canceling
$p_k$, we need the solutions of
\[ \ln \left(1 + \frac{\Pi}{\alpha_k}\right) \alpha_k \gamma_k -
\frac{\eta \Pi}{\alpha_k + \Pi} = 0 \] for $\Pi_{k-1} < \Pi \leq
\Pi_{k}$ when $1 \leq k \leq |{\cal H}| - 1$, and for $\Pi_{|{\cal H}|
  -1} < \Pi < \infty$ when $k = |{\cal H}|$, with the break-points
$\Pi_k, 1 \leq k \leq |{\cal H}|,$ as given earlier.  Let us write
$\frac{\Pi}{\alpha_k + \Pi} = 1 - \frac{1}{1 + \frac{\Pi}{\alpha_k}}$,
define $b_k = \ln \alpha_k \gamma_k$ (observe that $b_1 = 0)$, and,
for given $k$, use the new variable
\[ y = \frac{1}{1 + \frac{\Pi}{\alpha_k}} \]
Note that, for $0 < \Pi < \infty$, $ 1 > y > 0$.
Define $\delta_k = \frac{1}{1 +  \frac{\Pi_k}{\alpha_k}} $. Then
we seek the solutions of
\[ \ln\frac{1}{y} + b_k - \eta \left( 1 - y \right) = 0\] for $
\delta_{k} \leq y < \delta_{k-1} $, for each $k, 1 \leq k \leq |{\cal
  H}|$; note that $\delta_0 = 1$, and $\delta_{|{\cal H}|} = 0$. The
equations can be written more simply as
\[ e^{b_k - \eta} = y e^{-\eta y}, \] and are depicted in
Figure~\ref{fig:gamma_d_stationary_points}.
At this point we can conclude the following
\begin{theorem}
\label{thm:gamma_d_stationary_points}
There are at most $2 |{\cal H}| - 1$ stationary points of
  $\Gamma(d) d$ in $0 < d < \infty$.
\end{theorem}
\proof{
The result follows from the
  arguments just before the theorem statement, since each line
  $e^{(b_k - \eta)}$ , for $ 2 \leq k \leq |{\cal H}|$, has at most
  two intersections with $ye^{-\eta y}$, in $0 < y < 1$, and $e^{-
    \eta}$ } has only one such intersection.
\hfill \QED

\subsection{Fixed Transmission Time vs Fixed Packet Size}
\label{sec:mh_ftt_vs_fp}
In this section, we will formally establish that fixed transmission time
schemes are more throughput efficient compared to fixed packet size
schemes, for a given average power constraint.
We will prove this result in a general framework, without explicitly modelling the underlying
MAC, the power control schemes used or the channel fading distribution.

\noindent
\textbf{Data Transmission Model:} In a fixed transmission time scheme,
all data transmissions (with positive rate) are of
a fixed amount of time $T$, independent of the channel state $h$
and the power used.
Earlier, in our work (see Section~\ref{sec:fixed_trans_time}),
we assumed that, when the channel fade is poor (and hence $P(h) = 0$), the channel
is left idle for the next $T$ seconds.
Further, the optimal power control policy for such a system was
found to be a non-randomized policy, where a node
transmits with constant power $P(h)$ every time the channel is in state $h$
(see Section~\ref{sec:optimize_ph}).
Here, we will allow the possibility of the channel \emph{being relinquished
when bad} with a fixed time overhead $\leq T$.
We consider a spatio-temporal fading process with successive
transmitter-receiver pairs being selected by a distributed
multiaccess contention mechanism.
Hence, relinquishing the channel might improve throughput,
as successive fade levels might have little correlation.
The optimal policy for such a MAC could be a randomized
policy.
Hence, we will allow a randomized power control, i.e., for a channel state
$h$, the transmitter chooses a power $P_h$ according to some distribution.
In a fixed packet size scheme,
all data transmissions (with positive rate) carry
a fixed amount of data $L$ independent of the channel state $h$
and the power control used. Here as well, we will allow the possibility
of a randomized power control and the possibility of relinquishing the channel
with a fixed time overhead (when the channel fade is poor).

\noindent
\textbf{Optimality Criterion: }
The \emph{throughput} optimality of a data transmission scheme is established
either by comparing the energy required to send a certain amount of bits in
a given time or by comparing the amount of bits sent with a given amount
of energy in a given time. (We will discuss more about this
optimality criterion in Remark~\ref{rem:mh_optimality_criterion}).
We study a data transmission scheme
by considering two data transmissions of positive rates, in some arbitrary
channel states with gains $h_1$ and $h_2$ and with applied powers
$P_{h_1}$ and $P_{h_2}$. We do not make any assumption on
the probabilities of $h_1$ and $h_2$, and about the power control policy
which yields the powers $P_{h_1}$ and $P_{h_2}$.


For a given power control scheme ($h, P_h$),
we will then assume that the transmission rate given by Shannon's formula
is achieved over the transmission burst;
i.e.,
the transmission rate is given by
\[ C_{h} = W \log(1 + h P_h) \]
We have absorbed the factor $\frac{\alpha}{\sigma^2 d^{\eta}}$ in to
the term $h$ (since $d$ is fixed in this discussion).
Hence, the time durations taken to transmit the $L$ bits during the channel states $h_1$ and
$h_2$ (with the powers $P_{h_1}$ and $P_{h_2}$) are given by
$T_{h_1} := \frac{L}{W \log(1 + h_1 P_{h_1})}$ and $T_{h_2} := \frac{L}{W \log(1 + h_2 P_{h_2})}$.
Then, the total time occupied by these two transmissions is
\begin{eqnarray}
\label{eqn:mh_fps_tp}
T_P = \frac{L}{W \log(1 + h_1 P_{h_1})} + \frac{L}{W \log(1 + h_2 P_{h_2})}
\end{eqnarray}
spending an amount of energy equal to
\begin{equation}
\label{eqn:mh_fps_ep}
E_P = \frac{L P_{h_1}}{W \log(1 + h_1 P_{h_1})} + \frac{L P_{h_2}}{W \log(1 + h_2 P_{h_2})}
\end{equation}
Define $L_P := 2 \times L$ as the amount of bits sent in time $T_P$ using
an energy $E_P$ in channel states $h_1$ and $h_2$.

\begin{lemma}
\label{lem:mh_fps_fading_rate}
Let $h_1 > h_2$.
For a fixed packet size scheme,
if $P_{h_1}$ and $P_{h_2}$ are applied powers during channel states
$h_1$ and $h_2$,
then having $h_1 P_{h_1} \geq h_2 P_{h_2}$ is
throughput optimal.
\end{lemma}
\begin{proof}
Suppose that $h_1 P_{h_1} < h_2 P_{h_2}$.
Then,
\[ \log(1 + h_1 P_{h_1}) < \log(1 + h_2 P_{h_2}) \]
Find power controls $\tilde{P}_{h_1}$ and $\tilde{P}_{h_2}$ such that
\begin{eqnarray}
\label{eqn:mh_fps_hph_1}
\log(1 + h_1 P_{h_1}) &=& \log(1 + h_2 \tilde{P}_{h_2}) \\
\label{eqn:mh_fps_hph_2}
\log(1 + h_2 P_{h_2}) &=& \log(1 + h_1 \tilde{P}_{h_1})
\end{eqnarray}
or, equivalently,
\begin{eqnarray}
\label{eqn:mh_newpower_1}
h_1 P_{h_1} &=& h_2 \tilde{P}_{h_2} \\
\label{eqn:mh_newpower_2}
h_2 P_{h_2} &=& h_1 \tilde{P}_{h_1}
\end{eqnarray}
With the power control scheme $(h_1, \tilde{P}_{h_1}), (h_2,\tilde{P}_{h_2})$, the total time occupied
in the transmissions of $2 \times L$ bits during the channel
states $h_1$ and $h_2$ is,
\begin{eqnarray*}
T_{\tilde{P}} &=& \frac{L}{W \log(1 + h_1 \tilde{P}_{h_1})} + \frac{L}{W \log(1 + h_2 \tilde{P}_{h_2})} \\
&=& T_P
\end{eqnarray*}
(from (\ref{eqn:mh_fps_hph_1}) and (\ref{eqn:mh_fps_hph_2})).
Now, consider the energy spent to transmit these $2 \times L$ bits, i.e.,
\begin{eqnarray*}
E_{\tilde{P}} = \frac{L \tilde{P}_{h_1}}{W \log(1 + h_1 \tilde{P}_{h_1})} +
\frac{L \tilde{P}_{h_2}}{W \log(1 + h_2 \tilde{P}_{h_2})}
\end{eqnarray*}
Substituting for $\tilde{P}_{h_1}$ and $\tilde{P}_{h_2}$ from (\ref{eqn:mh_newpower_1})
and (\ref{eqn:mh_newpower_2}), we have,
\begin{eqnarray*}
E_{\tilde{P}} =
\frac{1}{h_1} \frac{L h_2 P_{h_2}}{W \log(1 + h_2 P_{h_2})} +
\frac{1}{h_2} \frac{L h_1 P_{h_1}}{W \log(1 + h_1 P_{h_1})}
\end{eqnarray*}
Rearranging the terms, we have,
\begin{eqnarray*}
E_{\tilde{P}} &=& \frac{1}{h_2} \frac{L h_1 P_{h_1}}{W \log(1 + h_1 P_{h_1})}
+ \frac{1}{h_1} \frac{L h_2 P_{h_2}}{W \log(1 + h_2 P_{h_2})} \\
&<& \frac{1}{h_1} \frac{L h_1 P_{h_1}}{W \log(1 + h_1 P_{h_1})}
+ \frac{1}{h_2} \frac{L h_2 P_{h_2}}{W \log(1 + h_2 P_{h_2})} \\
&=& \frac{L P_{h_1}}{W \log(1 + h_1 P_{h_1})}
+ \frac{L P_{h_2}}{W \log(1 + h_2 P_{h_2})} \\
&=& E_{P}
\end{eqnarray*}
where the inequality follows from the fact that
\begin{eqnarray*}
\lefteqn{\frac{L h_1 P_{h_1}}{W \log(1 + h_1 P_{h_1})} \left( \frac{1}{h_2} - \frac{1}{h_1} \right)} \\
&& < \frac{L h_2 P_{h_2}}{W \log(1 + h_2 P_{h_2})} \left( \frac{1}{h_2} - \frac{1}{h_1} \right)
\end{eqnarray*}
since $h_1 > h_2$ and $h_1 P_{h_1} < h_2 P_{h_2}$ (by assumption)
and the fact that $\frac{x}{\log(1 + x)}$ is strictly monotone increasing.

It follows that an optimal power control must have
$h_1 P_{h_1} \geq h_2 P_{h_2}$.
\end{proof}

\begin{remark}
From Lemma~\ref{lem:mh_fps_fading_rate}, we see that, when $h_1 > h_2$,
$C_{h_1} := W \log(1 + h_1 P_{h_1}) \geq W \log(1 + h_2 P_{h_2}) =: C_{h_2}$,
or equivalently, $T_{h_1} \leq T_{h_2}$.  \hfill \QED
\end{remark}

We will now provide a comparison of the fixed packet scheme with a
fixed transmission time scheme and show the optimality of the
fixed transmission time schemes.
The comparison
is done under the following assumption.
\begin{itemize}
\item The channel has the same marginal fading distribution,
whenever sampled by a transmitter, for either schemes. This is a reasonable
assumption as we consider spatio-temporal fading, with successive transmissions
from possibly different source-destination pairs chosen by the distributed
multiaccess contention scheme.
\end{itemize}
For the fixed packet size scheme, $L_P := 2 \times L$ bits were transmitted
in $T_P (= T_{h_1} + T_{h_2})$ time (see (\ref{eqn:mh_fps_tp}))
with an amount of energy equal
to $E_P$ (see (\ref{eqn:mh_fps_ep})), in two channel samples $h_1$ and $h_2$.
A reasonable comparison would be to
find the throughput of a fixed transmission time scheme for a
total duration of $T_P$ seconds
involving two data transmissions
with channel samples $h_1$ and $h_2$
of equal duration $T = \frac{T_P}{2}$ and a total energy of $E_P$.
We will assume that $P_{h_1}$ and $P_{h_2}$, the power used for
the fixed packet size scheme are such that $T_{h_1} \leq T_{h_2}$
(see Lemma~\ref{lem:mh_fps_fading_rate}).
Hence, we have $T_{h_1} \leq T \leq T_{h_2}$,
or, the fixed transmission time scheme spends relatively more time
on a better channel.
Clearly, its throughput is better than the fixed
packet size scheme for the same energy constraint, as seen below.

Let $P_{t_{h_1}}$ and $P_{t_{h_2}}$ be the optimal power control
for the fixed transmission time strategy such that
\[ E_T := P_{t_{h_1}} T + P_{t_{h_2}} T =  P_{h_1} T_{h_1} + P_{h_2} T_{h_2} = E_P\]
We have,
\begin{eqnarray*}
L_P = 2 L = T_{h_1} W \log(1 + h_1 P_{h_1}) + T_{h_2} W \log(1 + h_2 P_{h_2})
\end{eqnarray*}
Expanding the left hand side, we have,
\begin{eqnarray*}
2 L &=& T_{h_1} W \log(1 + h_1 P_{h_1}) + (T_{h_2} - T) W \log(1 + h_2 P_{h_2}) \\
&+& T W \log(1 + h_2 P_{h_2})
\end{eqnarray*}
Using $h_1 > h_2$, we get,
\begin{eqnarray*}
2 L &\leq& T_{h_1} \log(1 + h_1 P_{h_1}) + (T_{h_2} - T) \log(1 + h_1 P_{h_2}) \\
&+& T \log(1 + h_2 P_{h_2}) \\
&\leq& T \log(1 + h_1 P_{t_{h_1}}) + T \log(1 + h_2 P_{t_{h_2}}) \\
&=:& L_T
\end{eqnarray*}
where the last inequality follows from the fact that
$(h_1, P_{t_{h_1}})$ and $(h_2, P_{t_{h_2}})$
is the optimal power control scheme for the fixed transmission time scheme
with time $T_P (= 2 \times T)$ and energy $E_T (= E_P)$.

\begin{remarks}
\label{rem:mh_optimality_criterion}
For $L(t)$ defined as the amount of bits sent up to time $t$, and $E(t)$
defined as the total energy spent up to time $t$,
the average throughput ($\Theta$) and the average power ($\bar{P}$)
of the system are, in general, defined as
\[ \Theta := \liminf_{t \rightarrow \infty} \frac{L(t)}{t} \]
\[ \bar{P} := \limsup_{t \rightarrow \infty} \frac{E(t)}{t} \]
Under additional assumptions on the fading process and the
power control scheme used, the expressions are simplified
as an ensemble average (for example, see (\ref{eqn:phi_case2}) and (\ref{eqn:pow_case2})
for a fixed transmission time scheme).
In this section, the optimality of the schemes have been shown directly,
by comparing the amount of bits transmitted
for a particular sample of channel
for a given amount of time and energy, or by comparing the amount
of energy used to transmit a given amount of bits for a particular sample of channel
in a given amount of time.
For example, the argument provided here
directly translates to an argument with the ensemble average for the discrete fading case.
This approach is not only straightforward, but
also is very general.  \hfill \QED
\end{remarks}

\end{document}